\documentclass[prb,twocolumn,showpacs,amsmath,amssymb]{revtex4}
\usepackage{graphicx}
\usepackage{texdraw}

\newsavebox{\cl}
\sbox{\cl}{
\begin{texdraw}
\move (0 0) \fcir f:0.9 r:0.028 \larc r:0.028 sd:0 ed:360
\end{texdraw}}

\newsavebox{\cld}
\sbox{\cld}{
\begin{texdraw}
\move (0 0) \fcir f:1.0 r:0.028 \larc r:0.028 sd:0 ed:360
\end{texdraw}}

\newsavebox{\clu}
\sbox{\clu}{
\begin{texdraw}
\move (0 0) \fcir f:0.0 r:0.028 \larc r:0.028 sd:0 ed:360
\end{texdraw}}

\newsavebox{\alll}
\sbox{\alll}{
\unitlength=0.40mm%
\begin{picture}(20,10)
\put(2.9,0){\usebox{\cl}}
\put(2.9,5.1){\usebox{\cl}}
\put(8,0){\usebox{\cl}}
\end{picture}}

\newsavebox{\ddd}
\sbox{\ddd}{
\unitlength=0.40mm%
\begin{picture}(9,5)
\put(0,0){\usebox{\cld}}
\put(0,5.1){\usebox{\cld}}
\put(5.1,0){\usebox{\cld}}
\end{picture}}

\newsavebox{\udd}
\sbox{\udd}{
\unitlength=0.40mm%
\begin{picture}(9,5)
\put(0,0){\usebox{\clu}}
\put(0,5.1){\usebox{\cld}}
\put(5.1,0){\usebox{\cld}}
\end{picture}}

\newsavebox{\dud}
\sbox{\dud}{
\unitlength=0.40mm%
\begin{picture}(9,5)
\put(0,0){\usebox{\cld}}
\put(0,5.1){\usebox{\clu}}
\put(5.1,0){\usebox{\cld}}
\end{picture}}

\newsavebox{\uud}
\sbox{\uud}{
\unitlength=0.40mm%
\begin{picture}(9,5)
\put(0,0){\usebox{\clu}}
\put(0,5.1){\usebox{\clu}}
\put(5.1,0){\usebox{\cld}}
\end{picture}}

\newsavebox{\duu}
\sbox{\duu}{
\unitlength=0.40mm%
\begin{picture}(9,5)
\put(0,0){\usebox{\cld}}
\put(0,5.1){\usebox{\clu}}
\put(5.1,0){\usebox{\clu}}
\end{picture}}

\newsavebox{\uuu}
\sbox{\uuu}{
\unitlength=0.40mm%
\begin{picture}(9,5)
\put(0,0){\usebox{\clu}}
\put(0,5.1){\usebox{\clu}}
\put(5.1,0){\usebox{\clu}}
\end{picture}}

\newsavebox{\sdddd}
\sbox{\sdddd}{
\unitlength=0.40mm%
\begin{picture}(9,5)
\put(0.7,1.0){$\diagup$}
\put(0.5,1.0){$\diagup$}
\put(0.3,1.0){$\diagup$}
\put(0,-1.1){\usebox{\cld}}
\put(0,4.0){\usebox{\cld}}
\put(5.1,-1.1){\usebox{\cld}}
\put(5.1,4.0){\usebox{\cld}}
\end{picture}}

\newsavebox{\suddd}
\sbox{\suddd}{
\unitlength=0.40mm%
\begin{picture}(9,5)
\put(0.7,1.0){$\diagup$}
\put(0.5,1.0){$\diagup$}
\put(0.3,1.0){$\diagup$}
\put(0,-1.1){\usebox{\clu}}
\put(0,4.0){\usebox{\cld}}
\put(5.1,-1.1){\usebox{\cld}}
\put(5.1,4.0){\usebox{\cld}}
\end{picture}}

\newsavebox{\sdudd}
\sbox{\sdudd}{
\unitlength=0.40mm%
\begin{picture}(9,5)
\put(0.7,1.0){$\diagup$}
\put(0.5,1.0){$\diagup$}
\put(0.3,1.0){$\diagup$}
\put(0,-1.1){\usebox{\cld}}
\put(0,4.0){\usebox{\clu}}
\put(5.1,-1.1){\usebox{\cld}}
\put(5.1,4.0){\usebox{\cld}}
\end{picture}}

\newsavebox{\suudd}
\sbox{\suudd}{
\unitlength=0.40mm%
\begin{picture}(9,5)
\put(0.7,1.0){$\diagup$}
\put(0.5,1.0){$\diagup$}
\put(0.3,1.0){$\diagup$}
\put(0,-1.1){\usebox{\clu}}
\put(0,4.0){\usebox{\clu}}
\put(5.1,-1.1){\usebox{\cld}}
\put(5.1,4.0){\usebox{\cld}}
\end{picture}}

\newsavebox{\suddu}
\sbox{\suddu}{
\unitlength=0.40mm%
\begin{picture}(9,5)
\put(0.7,1.0){$\diagup$}
\put(0.5,1.0){$\diagup$}
\put(0.3,1.0){$\diagup$}
\put(0,-1.1){\usebox{\clu}}
\put(0,4.0){\usebox{\cld}}
\put(5.1,-1.1){\usebox{\cld}}
\put(5.1,4.0){\usebox{\clu}}
\end{picture}}

\newsavebox{\sduud}
\sbox{\sduud}{
\unitlength=0.40mm%
\begin{picture}(9,5)
\put(0.7,1.0){$\diagup$}
\put(0.5,1.0){$\diagup$}
\put(0.3,1.0){$\diagup$}
\put(0,-1.1){\usebox{\cld}}
\put(0,4.0){\usebox{\clu}}
\put(5.1,-1.1){\usebox{\clu}}
\put(5.1,4.0){\usebox{\cld}}
\end{picture}}

\newsavebox{\sduuu}
\sbox{\sduuu}{
\unitlength=0.40mm%
\begin{picture}(9,5)
\put(0.7,1.0){$\diagup$}
\put(0.5,1.0){$\diagup$}
\put(0.3,1.0){$\diagup$}
\put(0,-1.1){\usebox{\cld}}
\put(0,4.0){\usebox{\clu}}
\put(5.1,-1.1){\usebox{\clu}}
\put(5.1,4.0){\usebox{\clu}}
\end{picture}}

\newsavebox{\suduu}
\sbox{\suduu}{
\unitlength=0.40mm%
\begin{picture}(9,5)
\put(0.7,1.0){$\diagup$}
\put(0.5,1.0){$\diagup$}
\put(0.3,1.0){$\diagup$}
\put(0,-1.1){\usebox{\clu}}
\put(0,4.0){\usebox{\cld}}
\put(5.1,-1.1){\usebox{\clu}}
\put(5.1,4.0){\usebox{\clu}}
\end{picture}}

\newsavebox{\suuuu}
\sbox{\suuuu}{
\unitlength=0.40mm%
\begin{picture}(9,5)
\put(0.7,1.0){$\diagup$}
\put(0.5,1.0){$\diagup$}
\put(0.3,1.0){$\diagup$}
\put(0,-1.1){\usebox{\clu}}
\put(0,4.0){\usebox{\clu}}
\put(5.1,-1.1){\usebox{\clu}}
\put(5.1,4.0){\usebox{\clu}}
\end{picture}}

\newsavebox{\dddd}
\sbox{\dddd}{
\unitlength=0.40mm%
\begin{picture}(9,5)
\put(0,-1.1){\usebox{\cld}}
\put(0,4.0){\usebox{\cld}}
\put(5.1,-1.1){\usebox{\cld}}
\put(5.1,4.0){\usebox{\cld}}
\end{picture}}

\newsavebox{\uddd}
\sbox{\uddd}{
\unitlength=0.40mm%
\begin{picture}(9,5)
\put(0,-1.1){\usebox{\clu}}
\put(0,4.0){\usebox{\cld}}
\put(5.1,-1.1){\usebox{\cld}}
\put(5.1,4.0){\usebox{\cld}}
\end{picture}}

\newsavebox{\uudd}
\sbox{\uudd}{
\unitlength=0.40mm%
\begin{picture}(9,5)
\put(0,-1.1){\usebox{\clu}}
\put(0,4.0){\usebox{\clu}}
\put(5.1,-1.1){\usebox{\cld}}
\put(5.1,4.0){\usebox{\cld}}
\end{picture}}

\newsavebox{\uddu}
\sbox{\uddu}{
\unitlength=0.40mm%
\begin{picture}(9,5)
\put(0,-1.1){\usebox{\clu}}
\put(0,4.0){\usebox{\cld}}
\put(5.1,-1.1){\usebox{\cld}}
\put(5.1,4.0){\usebox{\clu}}
\end{picture}}

\newsavebox{\duuu}
\sbox{\duuu}{
\unitlength=0.40mm%
\begin{picture}(9,5)
\put(0,-1.1){\usebox{\cld}}
\put(0,4.0){\usebox{\clu}}
\put(5.1,-1.1){\usebox{\clu}}
\put(5.1,4.0){\usebox{\clu}}
\end{picture}}

\newsavebox{\uuuu}
\sbox{\uuuu}{
\unitlength=0.40mm%
\begin{picture}(9,5)
\put(0,-1.1){\usebox{\clu}}
\put(0,4.0){\usebox{\clu}}
\put(5.1,-1.1){\usebox{\clu}}
\put(5.1,4.0){\usebox{\clu}}
\end{picture}}

\begin{document}

\title{Ground states of an Ising model on an extended Shastry-Sutherland lattice and
 the 1/2-magnetization plateau in some rare-earth-metal tetraborides}

\author{Yu.I. Dublenych}
\affiliation{Institute for Condensed Matter Physics, National
Academy of Sciences of Ukraine, 1 Svientsitskii Street, 79011
Lviv, Ukraine}
\date{\today}

\begin{abstract}{A complete solution of the ground-state problem
for an Ising model on the Shastry-Sutherland lattice with an
additional interaction along the diagonals of ``empty'' squares in
an applied magnetic field is presented. A rigorous proof is given
that this interaction gives rise to a plateau at one-half of the
saturation magnetization. Such a fractional plateau has been
observed in some rare-earth-metal tetraborides, in particular, in
strong Ising magnets ErB$_4$ (where it is the only one) and
TmB$_4$ (where it is the broadest one), but its origin has
remained unclear. Our study sheds new light on the solution of
this problem.}
\end{abstract}

\maketitle

\section{Introduction}

Here we consider an Ising model on the Shastry-Sutherland (SS)
lattice with an additional diagonal interaction. The SS lattice is
topologically equivalent to the Archimedean $3^2.4.3.4$ lattice
(Fig.~1). It has been shown that the ground-state problem of the
quantum Heisenberg model on the SS lattice in an applied magnetic
field can be solved exactly. \cite{bib1} This model has attracted
considerable interest when the magnetic subsystem of some
compounds has been shown to consist of weakly coupled layers of
magnetic ions arranged on a lattice that is topologically
equivalent to the SS one. These compounds are referred to as
Shastry-Sutherland magnets. SrCu${}_2$(BO${}_3$)$_2$ is the most
known and the most studied among them. \cite{bib2,bib3} But during
the last decade, many other SS magnets have been discovered. In
particular, this concerns an entire group of rare-earth-metal
tetraborides $R$B${}_4$ ($R$ = La -- Lu). Magnetic ions $R^{3+}$
carry spins large enough to be considered as classical Heisenberg
ones. If, in addition, the crystal field effects are strong, then
the magnets can be described in terms of the effective spin-1/2
model under strong Ising anisotropy. This is the case of TmB$_4$
and ErB$_4$.

SS magnets have attracted particular interest, since they exhibit
sequences of fractional magnetization plateaus. A single plateau
at one-half of the saturation magnetization ($m/m_s = 1/2$) has
been observed in ErB$_4$. TmB$_4$, in addition to an extended
1/2-plateau, exhibits a sequence of narrow plateaus with
fractional values of $m/m_s = 1/6, 1/7$ up to 1/12 for
temperatures below 4 K (the magnetic field is normal to SS
planes). \cite{bib4,bib5,bib6}

\begin{figure}[]
\begin{center}
\includegraphics[scale = 1.0]{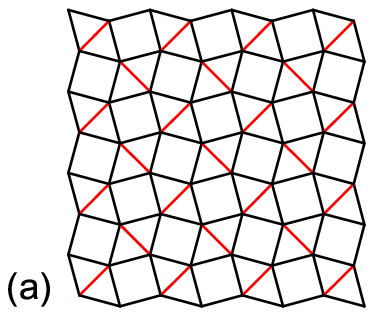}
\hspace{0.5cm}
\includegraphics[scale = 1.35]{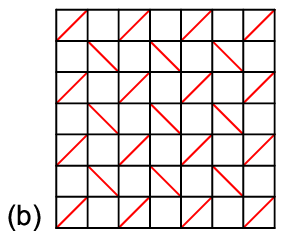}

\vspace{0.5cm}

\includegraphics[scale = 1.0]{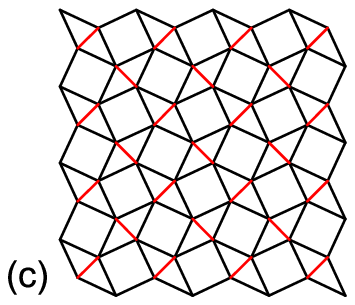}
\hspace{0.5cm}
\includegraphics[scale = 1.35]{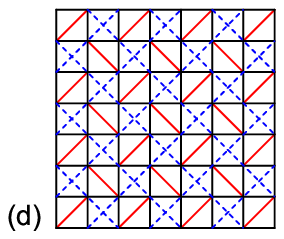}
\caption{(a) Archimedean $3^2.4.3.4$ lattice, (b)
Shastry-Sutherland lattice and (c) the lattice formed by magnetic
Cu$^{2+}$ ions in SrCu${}_2$(BO${}_3$)$_2$. All the three lattices
are topologically equivalent. (d) An extended Shastry-Sutherland
lattice considered here.}
\label{fig1}
\end{center}
\end{figure}

Some attempts have been made to explain the origin of the
fractional magnetization plateaus in terms of the Ising model on
the SS lattice, however, this model was shown, both numerically
\cite{bib7,bib8} and analytically, \cite{bib9} to predict a single
fractional plateau---at $m/m_s = 1/3$. This suggested that only
quantum models and maybe with longer-range interactions could
describe the 1/2-plateau and other fractional plateaus observed in
the rare-earth-metal tetraborides. \cite{bib7,bib10} In
Refs.~\onlinecite{bib10} and \onlinecite{bib11}, a 1/2-plateau was
obtained using the quantum SS model (and also its Ising limit)
with additional interactions $J_3$ (along the diagonals of
``empty'' squares) and $J_4$ (the next-nearest neighbors along
edges).

In our previous work, we showed that the fractional plateaus in an
Ising-type model on the SS lattice can be generated by the
long-range RKKY-type interactions (these are present in
rare-earth-metal tetraborides since the latter are good metals),
and that the 1/2-plateau is given rise by the additional diagonal
interaction $J_3$. \cite{bib9} Recently, a 1/2-plateau (as well as
a narrow 2/3-plateau) was shown numerically to be generated by
dipole-dipole interactions. \cite{bib12}

In our opinion, to explain the sequence of the magnetization
plateaus in TmB$_4$, it is sufficient to find the ground states of
an Ising-type model on an extended SS lattice with long-range
interactions. It is difficult to find a numerical solution but our
analytical method for determining the ground states of lattice gas
models or equivalent spin ones makes it possible.

Here, we present a complete solution of the ground-state problem
for the Ising model on the SS lattice with additional interaction
$J_3$ along the diagonals of ``empty'' squares (in what follows,
we refer to this model as the Ising model on the extended SS
lattice). To find this solution, we use the method of basic rays
and basic sets of cluster configurations which was developed in
our previous works. \cite{bib13,bib14} Moreover, we generalize the
method, and consider configurations of two (instead of one)
different clusters that is quite natural for the SS lattice with
additional diagonal bonds.

We rigorously prove the existence of a rather wide 1/2-plateau in
this model. It corresponds to three different phases and appears
under the condition that the interaction $J_1$ along the edges of
squares and at least one of the two remaining
interactions---along the SS diagonals ($J_2$) and along the
diagonals of the ``empty'' squares ($J_3$)---is
antiferromagnetic. It depends on the signs of the interactions
$J_2$ and $J_3$ which of these three phases is realized. A single
1/2-plateau exists, in particular, in the case when the
interactions $J_1$ and $J_2$ are antiferromagnetic and $J_3$ is
ferromagnetic (under the condition $J_2 < -2J_3$). A single
1/2-plateau has been observed in ErB$_4$, but its origin remains
unclear. Hence, our investigation sheds light on this problem.

The paper is organized as follows. In Sec.~II, a two-cluster
approach to the determination of the ground states of Ising-type
models is developed and a solution of the Ising model on the
extended SS lattice is found. In Sec.~III, the full-dimensional
structures are determined and analyzed. In Sec.~IV, the structures
at the three-dimensional boundaries of the full-dimensional
regions are considered and applied to predict the order of phase
transitions between full-dimensional phases. These are also
important for the investigation of the effects of longer-range
interactions. In Sec.~V, the ground states at the two-dimensional
boundaries of the full-dimensional regions are analyzed. In
Sec.~VI, the ground-state phase diagrams in the $(h, J_2)$-plane
are presented as well as possible sequences of phases for the
magnetization processes in  ErB$_4$ and TmB$_4$. Section VII gives
some conclusions.

\section{Solution of the ground-state problem for the
Ising model on the extended SS lattice: A two-cluster approach}

\begin{figure}[]
\begin{center}
\includegraphics[scale = 1.75]{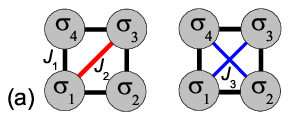}

\vspace{0.5cm}
\includegraphics[scale = 1.25]{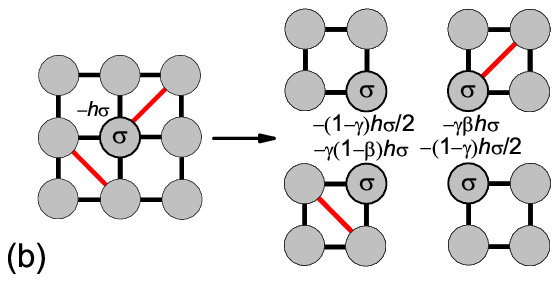}
\caption{(a) Clusters for the determination of the ground states:
(left) a cluster with an SS diagonal (interaction $J_2$) and
(right) a cluster without SS diagonal, but with ordinary ones
(interaction $J_3$). (b) The energy distribution of a site between
four squares sharing this site.}
\label{fig2}
\end{center}
\end{figure}

\begin{table*}
\caption{Basic rays and basic sets of configurations for the Ising
model on the extended Shastry-Sutherland lattice.}
\begin{ruledtabular}
\begin{tabular}{cclccc}
\multicolumn{2}{c}{Basic ray}&\multicolumn{1}{c}{Basic set}&Full-dimensional&``Free''\\\
$\mathbf{r}_i$&($h$, $J_1$, $J_2$, $J_3$)&\multicolumn{1}{c}{of configurations $\mathbf{R}_i$}&structures&coefficients\\
\hline\\[-2mm]
$\mathbf{r}_1$&$(0, 0, 0, -1)$&\usebox{\dddd} \usebox{\uddu} \usebox{\uuuu}&1, $\bar 1$, 3, 9, 10, $\overline {10}$&Arbitrary\\%[2mm]

$\mathbf{r}_2$&$(0, 0, -1, 0)$&\usebox{\sdddd} \usebox{\sdudd}
\usebox{\suddu} \usebox{\sduud} \usebox{\suduu} \usebox{\suuuu}
&1, $\bar 1$, 3, 7, 8&Arbitrary\\%[2mm]

$\mathbf{r}_3$&$(0, -1, 0, 1)$&\usebox{\dddd} \usebox{\uudd} \usebox{\uuuu}&1, $\bar 1$, 5, 8&$\alpha = 0$\\%[2mm]

$\mathbf{r}_4$&$(0, -1, 2, 0)$&\usebox{\sdddd} \usebox{\suddd}
\usebox{\suudd} \usebox{\sduuu} \usebox{\suuuu}
&1, $\bar 1$, 5, 9&$\alpha = 1$\\%[2mm]

$\mathbf{r}_5$&$(0, 1, 2, 0)$&\usebox{\suddd} \usebox{\suudd}
\usebox{\suddu} \usebox{\sduud} \usebox{\sduuu}
&3, 4, $\bar 4$, 5, 9&$\alpha = 1$\\%[2mm]

$\mathbf{r}_6$&$(2, 0, 0, 1)$&\usebox{\uudd} \usebox{\duuu}
\usebox{\uuuu}
&1, 5, 6, 7, 8&$\gamma = 0$\\%[2mm]

$\mathbf{r}_7$&$(4, 1, 0, 0)$&\usebox{\suddu} \usebox{\sduud}
\usebox{\sduuu} \usebox{\suduu} \usebox{\suuuu}
&1, 3, 4, 6, 7, 10&$\alpha = 1$, $\beta = \frac12$, $\gamma = 1$\\%[2mm]

$\mathbf{r}_8$&$(1, 0, 1, 0)$&\usebox{\suddd} \usebox{\suudd}
\usebox{\suddu} \usebox{\sduuu} \usebox{\suduu} \usebox{\suuuu}
&1, 4, 5, 6, 9, 10&$\beta = 1$, $\gamma = 1$\\%[2mm]

$\mathbf{r}_9$&$(2, 1, 0, 1)$&\usebox{\uudd} \usebox{\uddu}
\usebox{\duuu}
&3, 4, 5, 6, 7, 8&$\alpha = 0$, $\gamma = 0$\\%[2mm]

$\mathbf{r}_{10}$&$(2, 1, 2, -1)$&\usebox{\suudd} \usebox{\suddu}
\usebox{\sduud} \usebox{\sduuu}*
\usebox{\suduu}* $\parallel$ \usebox{\dddd} \usebox{\uddu} \usebox{\uuuu}
\usebox{\duuu}*\footnote{\mbox{Configurations marked by asterisk enter the structures in blocks shown in Fig.~4.}}
&3, 4, 9, 10&$\alpha = 1$, $\beta = \frac12$, $\gamma = 1$\\%[2mm]

$\mathbf{r}^{-}_6$&$(-2, 0, 0, 1)$&\usebox{\dddd} \usebox{\uddd}
\usebox{\uudd}
&$\bar 1$, 5, $\bar 6$, $\bar 7$, 8&$\gamma = 0$\\%[2mm]

$\mathbf{r}^{-}_7$&$(-4, 1, 0, 0)$&\usebox{\sdddd} \usebox{\suddd}
\usebox{\sdudd} \usebox{\suddu} \usebox{\sduud}
&$\bar 1$, 3, $\bar 4$, $\bar 6$, $\bar 7$, $\overline {10}$&$\alpha = 1$, $\beta = \frac12$, $\gamma = 1$\\%[2mm]

$\mathbf{r}^{-}_8$&$(-1, 0, 1, 0)$&\usebox{\sdddd} \usebox{\suddd}
\usebox{\sdudd} \usebox{\suudd} \usebox{\sduud} \usebox{\sduuu}
&$\bar 1$, $\bar 4$, 5, $\bar 6$, 9, $\overline {10}$&$\beta = 1$, $\gamma = 1$\\%[2mm]

$\mathbf{r}^{-}_9$&$(-2, 1, 0, 1)$&\usebox{\uudd} \usebox{\uddu}
\usebox{\uddd}
&3, $\bar 4$, 5, $\bar 6$, $\bar 7$, 8&$\alpha = 0$, $\gamma = 0$\\%[2mm]

$\mathbf{r}^{-}_{10}$&$(-2, 1, 2, -1)$&\usebox{\suudd}
\usebox{\suddu} \usebox{\sduud} \usebox{\suddd}* \usebox{\sdudd}*
$\parallel$
\usebox{\dddd} \usebox{\uddu} \usebox{\uuuu} \usebox{\uddd}*\footnote{Condition symmetric to the condition for $\mathbf{r}_{10}$.}
&3, $\bar 4$, 9, $\overline {10}$&$\alpha = 1$, $\beta = \frac12$,
$\gamma = 1$
\label{table1}
\end{tabular}
\end{ruledtabular}
\end{table*}

In our previous studies, \cite{bib9,bib13,bib14} we have
constructed the ground-state structures for Ising-type models (or
equivalent lattice-gas models) with configurations of some
cluster. We used only one kind of cluster. For instance, a cluster
in the form of a triangle with an SS diagonal as the hypotenuse is
sufficient for the Ising model on the conventional SS lattice.
\cite{bib9} Here we consider the same model but with additional
interaction $J_3$ along all the diagonals of the squares without
SS bonds [Fig.~1(d)]. To construct the ground-state structures for
this model one small cluster is insufficient; one should use two
different clusters: a square with an SS diagonal and a square
without it [Fig.~2(a)]. However, there is no crucial difference in
comparison with the one-cluster approach. The main idea is the
same and consists in introducing ``free'' coefficients which
account for the fact that the energy contribution of sites and
some bonds can be distributed between clusters in various ways.
(As will be clear later, ``free'' coefficients make it possible to
equalize and minimize the energies of two or more cluster
configurations at once). Each edge belongs to two squares (one of
each type) and each site belongs to four squares (two of each
type). The energy contribution of an edge (site) can be
distributed in various ways between the squares sharing this edge
(site). The way of distribution is determined by the ``free''
coefficients  $\alpha$, $\beta$, and $\gamma$ in the expressions
for the energy contributions of  different squares [see Fig.~2(b)]
\begin{eqnarray}
&&e = J_2\sigma_1\sigma_3 + \alpha J_1(\sigma_1\sigma_2 +
\sigma_2\sigma_3 + \sigma_3\sigma_4 + \sigma_4\sigma_1)\nonumber\\
&&~~~-\gamma h[\beta(\sigma_1 + \sigma_3) + (1 - \beta)(\sigma_2 + \sigma_4)],\nonumber\\
&&\tilde e = J_3(\sigma_1\sigma_3 + \sigma_2\sigma_4) +
(1-\alpha)J_1(\sigma_1\sigma_2 + \sigma_2\sigma_3 \nonumber\\
&&~~~+ \sigma_3\sigma_4 +
\sigma_4\sigma_1)-\frac{1-\gamma}{2}h(\sigma_1 + \sigma_2 +
\sigma_3 + \sigma_4).
\label{eq1}
\end{eqnarray}
Here $e$ and $\tilde e$ are the energy contributions of the
squares with SS bond and without SS bond, respectively; $J_1$,
$J_2$, and $J_3$ are the parameters of interaction along the
edges, SS diagonals, and ordinary diagonals (only for the squares
without SS diagonals), respectively; $h$ is the applied magnetic
field. The numbers $\sigma_1$, $\sigma_2$, $\sigma_3$, and
$\sigma_4$ ($\sigma_i = \pm 1$) define the spin configuration of a
square. There are nine configurations of a square with an SS
diagonal and six configurations of a square without SS diagonal:
\usebox{\sdddd}, \usebox{\suddd}, \usebox{\sdudd}, \usebox{\suudd},
\usebox{\suddu}, \usebox{\sduud}, \usebox{\sduuu}, \usebox{\suduu},
\usebox{\suuuu} $\parallel$ \usebox{\dddd}, \usebox{\uddd}, \usebox{\uudd},
\usebox{\uddu}, \usebox{\duuu}, \usebox{\uuuu}
[open and solid circles denote spin down ($-1$) and spin up
($+1$), respectively; the squares of different types are separated
by the symbol $\parallel$].

As in our previous studies, here we use the method of basic rays
(vectors) and basic sets of cluster configurations. In brief it is
as follows.

Consider the parameter space $(h, J_1, J_2, J_3)$ of the model.
Each ground-state structure corresponds to a region in this space.
The region can be presented as a solution of a set of uniform
linear inequalities; therefore, it is a polyhedral cone. Let us
recall that a polyhedral cone is a conical hull---that is, all
the linear combinations with nonnegative coefficients---of a set
of vectors. It is completely determined by its edges or vectors
along them. Among the polyhedral cones corresponding to all
possible ground-state structures of the model, the most important
are those with the dimensionality equal to the dimensionality of
the parameter space. We refer to these regions and corresponding
structures as full-dimensional.

If a structure is a ground-state one at two points of the
parameter space, then it is a ground-state structure in the entire
line segment connecting these points. This is a convexity
property; the region corresponding to a structure is always
convex. The convexity property makes it possible to determine the
ground-state structures in any point of the parameter space if all
edges of the full-dimensional regions (basic rays) and all
ground-state structures in these edges are known.

First we determine the cluster configurations which generate
ground-state structures and then the structures themselves. By
definition, a set of cluster configurations generates a structure
if all the cluster configurations in the structure belong to the
set. A set of cluster configurations generates all the structures
at the point $(h, J_1, J_2, J_3)$ of the parameter space if the
following three conditions are satisfied: (1) values of the
``free'' coefficients $\alpha$, $\beta$, and $\gamma$ in
Eqs.~(\ref{eq1}) can be chosen in such a manner that all the
configurations of the set have the same energy, (2) this energy is
lower than the energies of all the remaining configurations, and
(3) at least one structure can be generated by the configurations
of the set. It should be noticed that this group of conditions is
sufficient but not necessary; only the third condition is
necessary.

The basic rays and the corresponding sets of cluster
configurations (basic sets) of the model under consideration are
listed in Table~I. Configurations of clusters of different types
are separated by the symbol $\parallel$. If configurations of a
type of cluster are not indicated in a set, then all the
configurations of this type belong to the set (i.e., the
configuration of this type of cluster can be arbitrary). The last
column of Table ~I contains the values of the ``free''
coefficients for which the configurations belonging to the basic
set have the same energy which is lower than the energies of the
remaining configurations (in the corresponding ray, of course).

Having determined the full-dimensional regions and structures on
the basis of Table~I, we can show that the set of basic rays is
complete. This will be done later. It should be noticed that we
present here only the final results, the correctness of which can
be easily verified. In reality, first we found the
full-dimensional ground-state structures of the model (all these
but one are determined in Ref.~\onlinecite{bib9}) and only then we
calculated the basic rays.

In all the basic rays, except for $\mathbf{r}_{10}$ and
$\mathbf{r}^{-}_{10}$, all the structures generated by the
configurations of the corresponding basic set are ground-state
structures in these rays. As to the rays $\mathbf{r}_{10}$ and
$\mathbf{r}^{-}_{10}$, an additional condition should be
satisfied. For the ray $\mathbf{r}_{10}$ this condition can be
formulated as follows: the configurations marked by the asterisk
enter the ground-state structures in the blocks shown in Fig.~4
and these blocks do not overlap either with themselves or with
other squares. (For the symmetric ray $\mathbf{r}^{-}_{10}$ the
symmetric condition should be satisfied). Let us prove this.

Using Eq.~(\ref{eq1}), one can calculate the energies of all the
configurations of the clusters at the point $h = 2J_1$, $J_2 =
2J_1$, and $J_3 = -J_1$ (the ray $\mathbf{r}_{10}$) for the
following values of the ``free'' coefficients: $\alpha = 1$,
$\beta = 1/2$, and $\gamma = 1$.

\noindent\usebox{\sduuu} ($-2J_1$); \usebox{\suudd},
\usebox{\suddu}, \usebox{\sduud} (0); \usebox{\suddd},
\usebox{\suduu} ($2J_1$); \usebox{\suuuu} ($4J_1$);
\usebox{\sdudd} ($6J_1$); \usebox{\sdddd} ($12J_1$);
\usebox{\dddd}, \usebox{\uddu}, \usebox{\uuuu} (0);
\usebox{\uddd}, \usebox{\duuu} ($2J_1$); \usebox{\uudd} ($4J_1$).

\noindent The zero energy level is shifted here by $2J_1 > 0$. The
energy value indicated after each group of configurations is the
same for all the members of the group. We refer to the squares
with negative, zero, and positive energies as ``negative,''
``zero,'' and ``positive'' squares, respectively.

Let us prove that, in the ray $\mathbf{r}_{10}$, the energy of an
arbitrary structure cannot be negative. Near each negative square
\usebox{\sduuu}, in the region shown in Fig.~3(a), at
least one square among the three squares of the region is
positive. Let us group all the squares in a structure in the
following way: each zero square forms a group by itself; each
negative square enters the group of the positive square situated
near it in the region shown in Fig.~3(a). If a negative square
enters two or three groups, then its energy is distributed into
equal parts between these groups. It it easy to see that, for such
a grouping rule, positive squares \usebox{\suddd},
\usebox{\sdddd}, and \usebox{\uddd} cannot form a group with any negative square.
Squares \usebox{\sdudd} and \usebox{\uudd} can form groups with
one negative square only, but the energy of the groups is
positive. Square \usebox{\suuuu} can form a zero-energy group with
two negative squares. In this case, however, other groups with
positive energies are formed inevitably [Fig.~3(b)]. The square
\usebox{\suduu} can form a sigle zero-energy group with a negative
square in the way shown in Fig.~3(b). The only square that can
form a group with negative energy ($-J_1$) is the square
\usebox{\duuu} [Fig.~3(d)]. However, then there appears one of the
two positive squares, \usebox{\uudd} or \usebox{\duuu}. The first
one forms a group with the energy $3J_1$ which cannot be
compensated by the negative energy of the group formed by the
square \usebox{\duuu}. The second square \usebox{\duuu} can form a
zero-energy group, but in this case the situation is repeated and
so the negative-energy group of the first square
\usebox{\duuu} generates an infinite half-stripe in which all the
remaining groups have zero energies. Hence, the number of
negative-energy groups in the structure [Fig.~3(d)] can be
infinitesimal only. To summarize, the ground-state structures in
the ray $\mathbf{r}_{10}$, can contain, along with zero squares,
the positive squares \usebox{\suduu} and \usebox{\duuu} but only
in combinations with the square \usebox{\sduuu} shown in Fig.~4.

\begin{figure}[]
\begin{center}
\includegraphics[scale = 1.25]{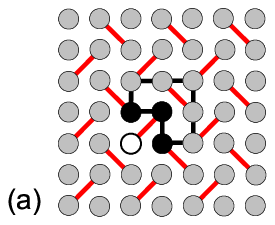}
\hspace{0.5cm}
\includegraphics[scale = 1.25]{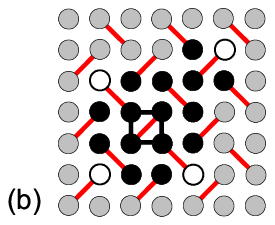}

\vspace{0.5cm}

\includegraphics[scale = 1.25]{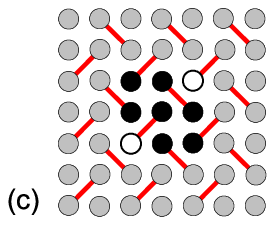}
\hspace{0.5cm}
\includegraphics[scale = 1.25]{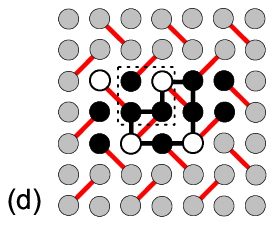}
\caption{(a) The region (solid line) near a negative square
containing at least one positive square. (b) Square
\usebox{\suuuu} can form a zero-energy group but some
positive-energy groups emerge then inevitably. (c) Zero-energy
group with a square \usebox{\suduu}. (d) Square \usebox{\duuu} can
form a negative-energy group (solid line), but in this case a new
square \usebox{\duuu} (dotted line) emerges with a group that
should have zero energy and, therefore, this square yields the
same square and so on.}
\label{fig3}
\end{center}
\end{figure}

\begin{figure}[]
\begin{center}
\includegraphics[scale = 1.5]{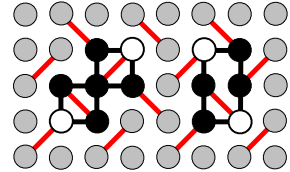}%{blokya.eps}
\caption{Blocks with the squares \usebox{\suduu} and
\usebox{\duuu} for the ground-state structures in the ray
$\mathbf{r}_{10}$.}
\label{fig4}
\end{center}
\end{figure}

\section{Full-dimensional ground-state structures and regions}

Table~I represents a complete solution of the ground-state problem
under consideration. Using this Table and the convexity property,
it is easy to determine the full-dimensional ground-state
structures and regions. All one has to do is to find the subsets
of the basic sets of configurations with the following properties:
(1) each of them is a subset of at least four basic sets and the
linear hull of the corresponding basic vectors is
full-(four-)dimensional; (2) the cluster configurations of such a
subset generate at least one structure.

\begin{table*}
\caption{Full-dimensional regions and ground-state structures of
the Ising model on the extended Shastry-Sutherland lattice (for $h
\geqslant 0$).}
\begin{ruledtabular}
\begin{tabular}{cllclcc}
Region&\multicolumn{1}{c}{Generating}&\multicolumn{1}{c}{Characteristics}&Magneti-&\multicolumn{1}{c}{}&Number&Condition for existence\\
(Structure)&\multicolumn{1}{c}{configurations}&\multicolumn{1}{c}{of the structure}&zation&\multicolumn{1}{c}{Basic rays}&of 3-faces&in the plane $(h, J_2)$\\
\hline\\[-2mm]
1&~~ \usebox{\suuuu} $\parallel$
\usebox{\uuuu}&$2J_1+\frac{J_2}{2}+J_3-h$&1& $\mathbf{r}_1,
\mathbf{r}_2, \mathbf{r}_3, \mathbf{r}_4,
\mathbf{r}_{6},$&8&Always\\
&&$[1 \parallel 1]$&&$\mathbf{r}_{7}, \mathbf{r}_{8}$&&\\[2mm]

3&~~ \usebox{\suddu} \usebox{\sduud} $\parallel$
\usebox{\uddu}&$-2J_1+\frac{J_2}{2}+J_3$&0& $\mathbf{r}_1, \mathbf{r}_2, \mathbf{r}_5,
\mathbf{r}_7, \mathbf{r}_9,$&11&$J_1 \geqslant 0$, $J_3 \leqslant J_1$\\
&&$\left[\frac12, \frac12 \parallel 1\right]$&&$ \mathbf{r}_{10}, \mathbf{r}_{7}^{-}, \mathbf{r}_{9}^{-}, \mathbf{r}_{10}^{-}$&&\\[2mm]

4&~~ \usebox{\suddu} \usebox{\sduuu} $\parallel$ \usebox{\uddu}
\usebox{\duuu}&$-\frac{2J_1}{3}-\frac{J_2}{6}+\frac{J_3}{3}-\frac{h}{3}$
&1/3&$\mathbf{r}_5, \mathbf{r}_7, \mathbf{r}_8, \mathbf{r}_9, \mathbf{r}_{10}$
&5&$J_2 \geqslant \max(0, -2J_3)$,\\
&&$\left[\frac13, \frac23 \parallel \frac13, \frac23\right]$&&&&$|J_3| \leqslant J_1$\\[2mm]

5&~~ \usebox{\suudd} $\parallel$
\usebox{\uudd}&$-\frac{J_2}{2}-J_3$&0& $\mathbf{r}_3, \mathbf{r}_4, \mathbf{r}_5,
\mathbf{r}_6, \mathbf{r}_8, \mathbf{r}_9$,&9&$J_3 \geqslant 0$\\
&&$\left[1 \parallel 1\right]$&&$\mathbf{r}_{6}^{-}, \mathbf{r}_{8}^{-}, \mathbf{r}_{9}^{-}$&&\\[2mm]

6&~~ \usebox{\suddu} \usebox{\sduuu} \usebox{\suduu} \usebox{\suuuu} $\parallel$ \usebox{\duuu}&
$-\frac{h}{2}$&1/2&
$\mathbf{r}_6, \mathbf{r}_7, \mathbf{r}_8,\mathbf{r}_9$&4&$J_1 \geqslant 0$, $J_2 \geqslant 0$, $J_3 \geqslant 0$\\
&&$\left[0, \frac12, \frac12, 0 \parallel 1\right]$&&&&\\[2mm]

7&~~ \usebox{\suddu} \usebox{\sduud} \usebox{\suduu}
\usebox{\suuuu} $\parallel$ \usebox{\duuu}&
$\frac{J_2}{2}-\frac{h}{2}$&1/2&
$\mathbf{r}_2, \mathbf{r}_6, \mathbf{r}_7, \mathbf{r}_9$&4&$J_1 \geqslant 0$, $J_2 \leqslant 0$, $J_3 \geqslant 0$\\
&&$\left[\mbox{Desorder} \parallel 1\right]$&&&&\\[2mm]

8&~~ \usebox{\sdddd} \usebox{\suddu} \usebox{\sduud}
\usebox{\suuuu} $\parallel$ \usebox{\uudd}&
$\frac{J_2}{2}-J_3$&0&$\mathbf{r}_2, \mathbf{r}_3, \mathbf{r}_6,
\mathbf{r}_9, \mathbf{r}_{6}^{-}, \mathbf{r}_{9}^{-}$&6&$J_2 \leqslant 0$, $J_3  \geqslant |J_1|$\\
&&$\left[\frac14, \frac14, \frac14, \frac14 \parallel 1\right]$&&&&\\[2mm]

9&~~ \usebox{\suudd} $\parallel$ \usebox{\dddd} \usebox{\uddu}
\usebox{\uuuu}& $-\frac{J_2}{2}+J_3$&0& $\mathbf{r}_1,
\mathbf{r}_4, \mathbf{r}_5, \mathbf{r}_8,
\mathbf{r}_{10},$&8&$J_2 \geqslant 2|J_1|$, $J_3 \leqslant 0$\\
&&$\left[1 \parallel \frac14, \frac12, \frac14\right]$&&$\mathbf{r}_{8}^{-}, \mathbf{r}_{10}^{-}$&&\\[2mm]

10&~~ \usebox{\sduuu} \usebox{\suduu} $\parallel$ \usebox{\uddu} \usebox{\uuuu}&$J_3-\frac{h}{2}$&1/2&
$\mathbf{r}_1, \mathbf{r}_7, \mathbf{r}_8, \mathbf{r}_{10}$&4&$J_2 \geqslant 0$, $J_3 \leqslant 0$\\
&&$\left[\frac12, \frac12 \parallel \frac12, \frac12\right]$&&&&\\
\label{table2}
\end{tabular}
\end{ruledtabular}
\end{table*}

The full-dimensional regions and the corresponding subsets of
configurations which generate all the ground-state structures in
these regions are listed in Table II. The first column gives the
numbers of full-dimensional regions. Taking into account the
symmetry of the model with respect to the field inversion with
simultaneous flip of all spins, we indicate only the regions
(structures) with zero and positive magnetization. We denote a
region (structure) with negative magnetization by the same number
as the symmetric region (structure) but with a bar over the
number. The third column of Table II gives energies per site of
structures and, in the square brackets, the fractional contents of
cluster configurations in the structures (see
Ref.~\onlinecite{bib15}). If the fractional contents of
configurations in a structure can vary (region 7), then there is a
degeneracy marked by the word ``Disorder.'' (In region 6, there is
degeneracy as well, but the fractional contents of configurations
do not vary). The fourth, fifth, sixth, and seventh columns
indicate, respectively: magnetization of the structure, basic rays
that define the corresponding region, the number of
three-dimensional faces of the region, and the conditions for the
existence of the region in the $(h, J_2)$-plane.

The full-dimensional ground-state structures of the model under
consideration are depicted in Figs.~5 and 6. They are constructed
with the cluster configurations listed in Table~II. All the
full-dimensional structures, except for structures 6 and 7, are
fully determinate. In the regions 6 an 7, an infinite number of
the ground-state structures occur: phases 6 and 7 are disordered
(Fig.~6). It is easy to see that the disorder of phase 7 is
one-dimensional: it is ordered in one direction and disordered in
another.

It is more difficult to determine the character (i.e., the
dimensionality) of disorder in phase 6. In structures of phase 6
(and even at the boundary between phases 4 and 6), the structural
element shown in Fig.~7 generates an infinite half-stripe. On the
basis of this fact one can prove that the disorder of phase 6 is
one-dimensional. The number of squares \usebox{\suddu} and
\usebox{\suuuu} is infinitezimal in comparison with the number of
remaining squares. Although their energies are equal to the
energies of other squares which generate structure 6, these
squares represent a kind of zero-energy defects. Hence, structure
6 can be considered as a simple mixture of structures $6a$ and
$6b$ depicted in Fig.~5.

\begin{figure*}[]
\begin{center}
\includegraphics[scale = 1.0]{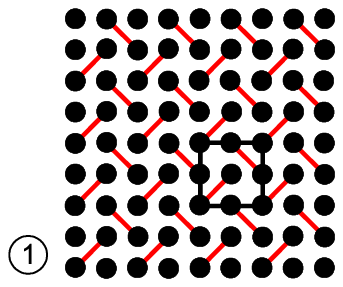}
\hspace{0.5cm}
\includegraphics[scale = 1.0]{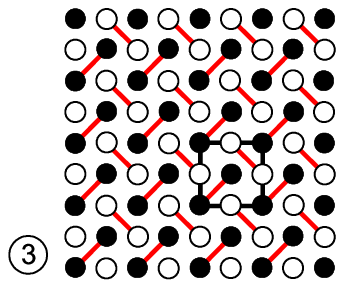}
\hspace{0.5cm}
\includegraphics[scale = 1.0]{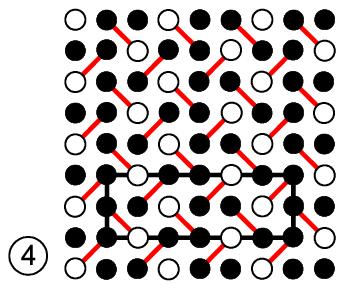}

\vspace{0.25cm}

\includegraphics[scale = 1.0]{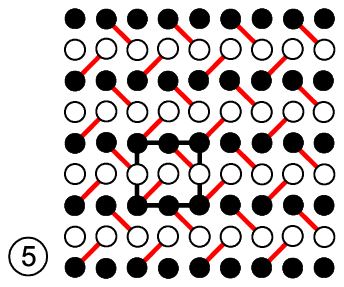}
\hspace{0.5cm}
\includegraphics[scale = 1.0]{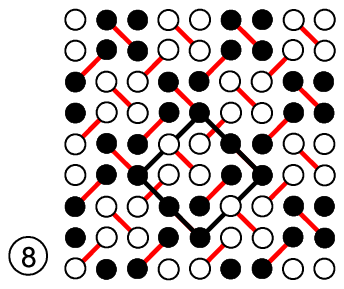}
\hspace{0.5cm}
\includegraphics[scale = 1.0]{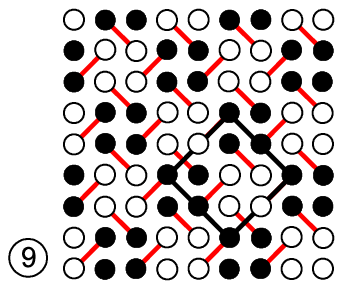}

\vspace{0.25cm}

\includegraphics[scale = 1.0]{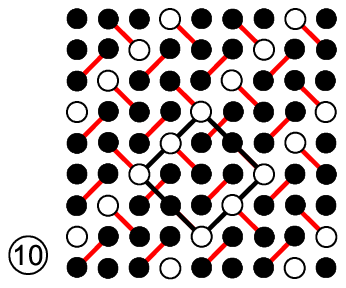}
\hspace{0.5cm}
\includegraphics[scale = 1.0]{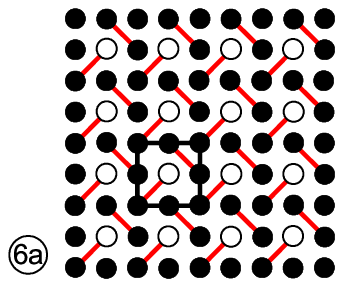}
\hspace{0.5cm}
\includegraphics[scale = 1.0]{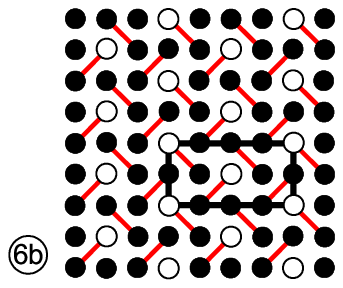}
\caption{Full-dimensional ground-state structures of the Ising
model on the extended SS lattice (for $h \geqslant 0$). Phases 1,
3, 4, and 5 are, respectively, the fully polarized phase, the
N\'{e}el phase, the 1/3-plateau phase (or the UUD phase), and the
collinear phase. Phases 8 and 9 are chessboard phases. Phases 10,
$6a$, and $6b$ are 1/2-plateau phases. A simple mixture of
structures $6a$ and $6b$ is the ground-state structure for region
6. Unit cells are indicated.}
\label{fig5}
\end{center}
\end{figure*}
\begin{figure*}[]
\begin{center}
\includegraphics[scale = 0.75]{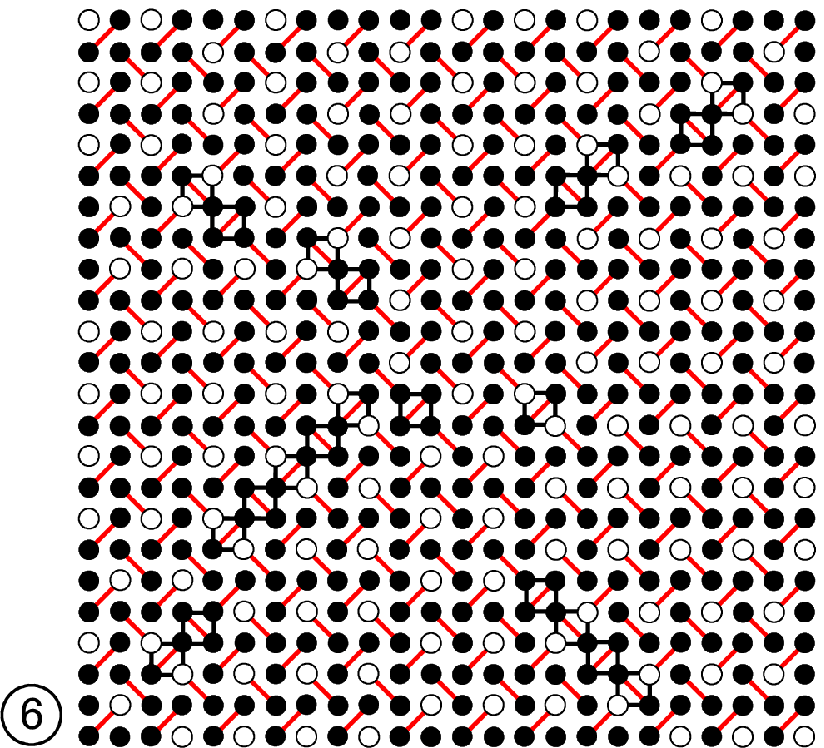}
\hspace{0.5cm}
\includegraphics[scale = 0.75]{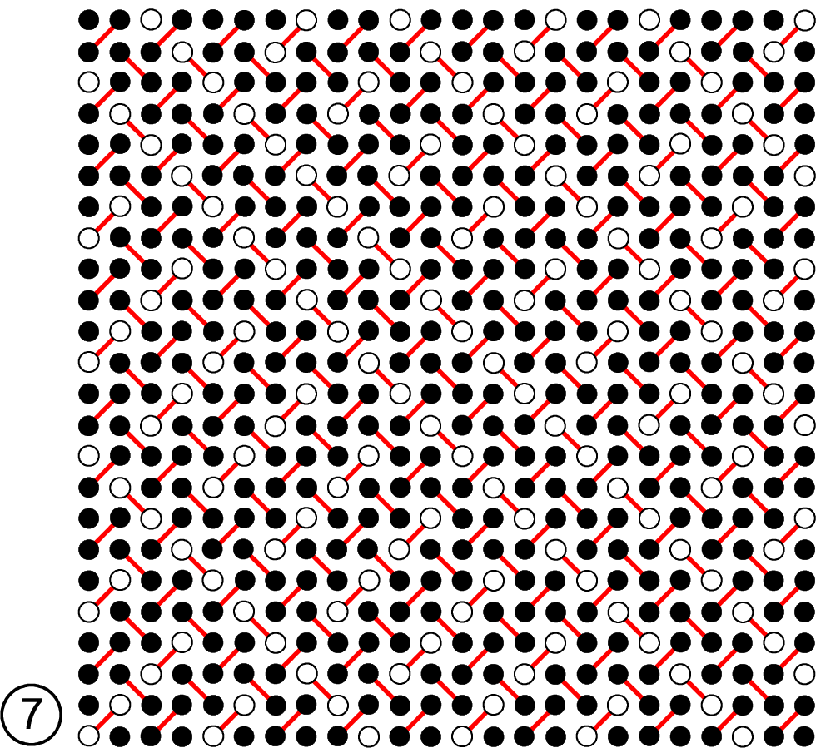}
\caption{ Full-dimensional ground-state structures of the Ising
model on the extended SS lattice. Disordered phases 6 and 7. In
structure 6 the squares-``defects'' are shown.}
\label{fig6}
\end{center}
\end{figure*}
\begin{figure}[t]
\begin{center}
\includegraphics[scale = 1.15]{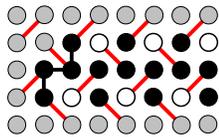}
\caption{A structural element in phases 4 and 6 (as well as in
their boundary) and an infinite half-stripe generated by this
element.}
\label{fig7}
\end{center}
\end{figure}
\begin{figure}[t]
\begin{center}
\includegraphics[scale = 1.25]{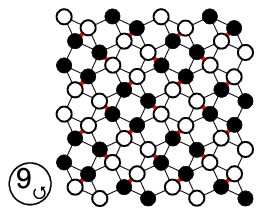}
\hspace{0.5cm}
\includegraphics[scale = 1.25]{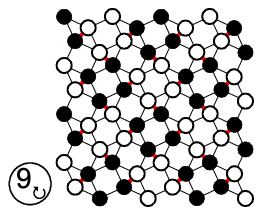}
\caption{Chirality of structure 9. The SS bonds twist
counterclockwise around ``open'' and ``solid'' squares (left
structure) or clockwise (right structure) but in the same manner
for all the squares.}
\label{fig8}
\end{center}
\end{figure}

It is interesting that the full-dimensional structure 9 is chiral.
This is better seen on the lattice shown in Fig.~1(c). Two
different structures 9 are possible: a structure with the SS bonds
twisted counterclockwise around ``open'' and ``solid'' squares and
its chiral twin with the SS bonds twisted clockwise (Fig.~8).
Hence, the interaction $J_3$ lifts the degeneracy of the Ising
dimer phase (opposite spins on the SS diagonals) \cite{bib9} and
thus two ordered phases are produced; one of these is chiral. It
should be noticed that this structure is not chiral on the
Archimedean lattice [Fig.~1(a)] with equal interactions $J_1$ and
$J_2$. Chiral structures are ubiquitous in nature, however, their
emergence as a result of spontaneous symmetry breaking remains
unclear. \cite{bib16}

\section{Ground-state structures in the 3-faces of the full-dimensional regions}

Let us consider now the ground-state structures in the
three-dimensional faces (3-faces) of the full-dimensional regions.
A 3-face is defined by a subset of the set of basic vectors for a
full-dimensional region. An algorithm for the determination of
3-faces of a full-dimensional region is described in
Ref.~\onlinecite{bib13}. It is a simple mathematical problem and
we do not reproduce the description of this algorithm here. Having
found the 3-faces of all the full-dimensional regions, one can
easily prove the completeness of the set of basic vectors. It is
sufficient to prove that each 3-face is a 3-face for two
full-dimensional regions and, hence, the full-dimensional regions
fill the whole parameter space without gaps and overlaps.

Knowing the basic vectors for a 3-face, it is easy to determine
the set of cluster configurations which generates all the
ground-state structures in this 3-face. This set is the
intersection of the basic sets of configurations for the basic
vectors of the 3-face. Then one can describe the corresponding
ground-state structures.

\begin{table*}
\caption{Basic rays and ground-state configurations for the
3-faces of the full-dimensional regions for the Ising model on the
extended SS lattice ($h \geqslant 0$). The dimensionality of
disorder is indicated for continuous transitions.}
\begin{ruledtabular}
\begin{tabular}{lllcc}
&\multicolumn{1}{c}{Basic rays}&\multicolumn{1}{c}{Ground-state configurations}&Transition&Conditions for existence\\
\multicolumn{1}{c}{Regions}&\multicolumn{1}{c}{of the 3-face}&\multicolumn{1}{c}{for the 3-face}&between phases&in the plane $(h, J_2)$\\
\hline\\[-2mm]
1, $\bar 1$&$\mathbf{r}_1, \mathbf{r}_2, \mathbf{r}_3, \mathbf{r}_4$&\usebox{\sdddd} \usebox{\suuuu} $\parallel$ \usebox{\dddd} \usebox{\uuuu}&Jump&$J_1 < 0$, $J_2 > \min[-2J_1, -2(J_1 + J_3)]$, $J_3 < |J_1|$\\
1, 3&$\mathbf{r}_1, \mathbf{r}_2, \mathbf{r}_7$&\usebox{\suddu} \usebox{\sduud} \usebox{\suuuu} $|$ \usebox{\suduu} $\parallel$ \usebox{\uddu} \usebox{\uuuu} $|$ \usebox{\dddd}&Jump&$J_1 > 0$, $J_2 < 0$, $J_3 < 0$\\
1, 5&$\mathbf{r}_3, \mathbf{r}_4, \mathbf{r}_8, \mathbf{r}_6$&\usebox{\suudd} \usebox{\suuuu} $|$ \usebox{\suddd} \usebox{\sduuu} $\parallel$ \usebox{\uudd} \usebox{\uuuu}&Cont. (1)&$J_1 < 0$, $J_2 > \max[0, -2(J_1 + J_3)]$, $J_3 > 0$\\
1, 6&$\mathbf{r}_6, \mathbf{r}_7, \mathbf{r}_8$&\usebox{\suddu} \usebox{\sduuu} \usebox{\suduu} \usebox{\suuuu} $\parallel$ \usebox{\duuu} \usebox{\uuuu} $|$ \usebox{\uudd}&Cont. (2)&$J_1 > 0$, $J_2 > 0$, $J_3 > 0$\\
1, 7&$\mathbf{r}_2, \mathbf{r}_6, \mathbf{r}_7$&\usebox{\suddu} \usebox{\sduud} \usebox{\suduu} \usebox{\suuuu} $\parallel$ \usebox{\duuu} \usebox{\uuuu} $|$ \usebox{\uudd}&Cont. (2)&$J_1 > 0$, $J_2 > 0$, $J_3 > 0$\\
1, 8&$\mathbf{r}_2, \mathbf{r}_3, \mathbf{r}_6$&\usebox{\sdddd} \usebox{\suddu} \usebox{\sduud} \usebox{\suduu} \usebox{\suuuu} $|$ \usebox{\sdudd} $\parallel$ \usebox{\uudd} \usebox{\uuuu}&Cont. (1)&$0 < -J_1 < J_3$, $J_2 < 0$\\
1, 9&$\mathbf{r}_1, \mathbf{r}_4, \mathbf{r}_8$&\usebox{\suudd} \usebox{\suuuu} $|$ \usebox{\suddd} \usebox{\sduuu} $\parallel$ \usebox{\dddd} \usebox{\uddu} \usebox{\uuuu}&Jump&$J_1 < 0$, $J_2 > -2J_1$, $J_3 < 0$\\
1, 10&$\mathbf{r}_1, \mathbf{r}_7, \mathbf{r}_8$&\usebox{\sduuu} \usebox{\suduu} \usebox{\suuuu} $|$ \usebox{\suddu} $\parallel$ \usebox{\uddu} \usebox{\uuuu} $|$ \usebox{\dddd}&Cont. (1)&$J_1 > 0$, $J_2 > 0$, $J_3 < 0$\\[2mm]

3, 4&$\mathbf{r}_5, \mathbf{r}_9, \mathbf{r}_7, \mathbf{r}_{10}$&\usebox{\suddu} \usebox{\sduud} \usebox{\sduuu} $\parallel$ \usebox{\uddu} \usebox{\duuu}&Cont. (1)&$\max(0, -2J_3) < J_2 < \min[2J_1, 2(J_1 - J_3)]$,\\
&&&&$J_1 > 0$, $|J_3| < J_1$\\
3, 5&$\mathbf{r}_5, \mathbf{r}_9, \mathbf{r}_9^{-}$&\usebox{\suddd} \usebox{\suudd} \usebox{\suddu} \usebox{\sduud} \usebox{\sduuu} $\parallel$ \usebox{\uudd} \usebox{\uddu}&Cont. (2)&$J_2 = 2(J_1 - J_3)$, $0 < J_3 < J_1$\\
3, 7&$\mathbf{r}_2, \mathbf{r}_7,  \mathbf{r}_9$&\usebox{\suddu} \usebox{\sduud} \usebox{\suduu} \usebox{\suuuu} $\parallel$ \usebox{\uddu} \usebox{\duuu} $|$ \usebox{\uudd}&Cont. (2)&$J_2 < 0$, $0 < J_3 < J_1$\\
3, 8&$\mathbf{r}_2, \mathbf{r}_9, \mathbf{r}_9^{-}$&\usebox{\sdddd} \usebox{\sdudd} \usebox{\suddu} \usebox{\sduud} \usebox{\suduu} \usebox{\suuuu} $\parallel$ \usebox{\uudd} \usebox{\uddu}&Cont. (1)&$J_2 < 0$, $J_3 = J_1$\\
3, 9&$\mathbf{r}_1, \mathbf{r}_{10}, \mathbf{r}_5, \mathbf{r}_{10}^{-}$&\usebox{\suudd} \usebox{\suddu} \usebox{\sduud} $\parallel$ \usebox{\dddd} \usebox{\uddu} \usebox{\uuuu}&Jump&$J_1 > 0$, $J_2 > 2J_1$, $J_3 < 0$\\
3, 10&$\mathbf{r}_1, \mathbf{r}_7, \mathbf{r}_{10}$&\usebox{\suddu} \usebox{\sduud} \usebox{\sduuu} \usebox{\suduu} $\parallel$ \usebox{\uddu} \usebox{\uuuu} $|$ \usebox{\dddd}&Cont. (1)&$J_1 > 0$, $0 < J_2 < \min(2J_1,-2J_3)$, $J_3 < 0$\\[2mm]

4, 5&$\mathbf{r}_5, \mathbf{r}_8, \mathbf{r}_9$&\usebox{\suddd} \usebox{\suudd} \usebox{\suddu} \usebox{\sduuu} $\parallel$ \usebox{\uudd} \usebox{\uddu} \usebox{\duuu}&Cont. (?)&$J_2 > 2(J_1 - J_3)$, $0 < J_3 < J_1$\\
4, 6&$\mathbf{r}_7, \mathbf{r}_8, \mathbf{r}_9$&\usebox{\suddu} \usebox{\sduuu} \usebox{\suduu} \usebox{\suuuu} $\parallel$ \usebox{\uddu} \usebox{\duuu} $|$ \usebox{\uudd}&Cont. (1)&$J_2 > 0$, $0 < J_3 < J_1$\\
4, 9&$\mathbf{r}_5, \mathbf{r}_8, \mathbf{r}_{10}$&\usebox{\suudd}
\usebox{\suddu} \usebox{\sduuu}$^{*}$ $\parallel$ \usebox{\dddd}
\usebox{\uddu} \usebox{\uuuu} \usebox{\duuu}$^{*}
$\footnote{\mbox{Configurations marked by asterisk enter the structures in blocks shown in Fig.~4.}}&Jump + Cont. (1)&$J_2 > 2J_1$, $-J_1 < J_3 < 0$\\
4, 10&$\mathbf{r}_7, \mathbf{r}_8, \mathbf{r}_{10}$&\usebox{\suddu} \usebox{\sduuu}$^{*}$ \usebox{\suduu}$^{*}$ $\parallel$ \usebox{\uddu} \usebox{\uuuu} \usebox{\duuu}$^{*}$ $|$ \usebox{\dddd}&Jump&$J_2 > -2J_3$, $-J_1 < J_3 < 0$\\[2mm]

5, 6&$\mathbf{r}_6, \mathbf{r}_8, \mathbf{r}_9$&\usebox{\suddd} \usebox{\suudd} \usebox{\suddu} \usebox{\sduuu} \usebox{\suduu} \usebox{\suuuu} $\parallel$ \usebox{\uudd} \usebox{\duuu}&Cont. (2)&$0 < J_1 < J_3$, $J_2 > 0$\\
5, 8&$\mathbf{r}_3, \mathbf{r}_6, \mathbf{r}_9, \mathbf{r}_9^{-}, \mathbf{r}_6^{-}$&\usebox{\uudd}&Cont. (1)&$J_2 = 0$, $J_3 > |J_1|$\\
5, 9&$\mathbf{r}_4, \mathbf{r}_8, \mathbf{r}_5, \mathbf{r}_8^{-}$&\usebox{\suddd} \usebox{\suudd} \usebox{\sduuu}&Cont. (2)&$J_2 > 2|J_1|$, $J_3  = 0$\\[2mm]

6, 7&$\mathbf{r}_6, \mathbf{r}_7, \mathbf{r}_9$&\usebox{\suddu} \usebox{\sduud} \usebox{\sduuu} \usebox{\suduu} \usebox{\suuuu} $\parallel$ \usebox{\duuu} $|$ \usebox{\uudd} $=$ \usebox{\duuu}&Cont. (2)&$J_1 > 0$, $J_2 = 0$, $J_3 > 0$\\[2mm]

7, 8&$\mathbf{r}_2, \mathbf{r}_6, \mathbf{r}_9$&\usebox{\sdddd} \usebox{\sdudd} \usebox{\suddu} \usebox{\sduud} \usebox{\suduu} \usebox{\suuuu} $\parallel$ \usebox{\uudd} \usebox{\duuu}&Cont. (2)&$0 < J_1 < J_3$, $J_2 < 0$\\[2mm]

9, 10&$\mathbf{r}_1, \mathbf{r}_8,
\mathbf{r}_{10}$&\usebox{\suudd} \usebox{\sduuu} \usebox{\suduu}
$|$ \usebox{\suddu} $\parallel$ \usebox{\dddd} \usebox{\uddu}
\usebox{\uuuu}&Cont. (1)&$J_3 < -J_1 < 0$, $J_2 > 2J_1$
\label{table3}
\end{tabular}
\end{ruledtabular}
\end{table*}

Basic vectors for the 3-faces of full-dimensional regions as well
as corresponding sets of cluster configurations are presented in
Table~III. In this Table, a 3-face is denoted by the numbers of
full-dimensional regions (phases) which share this 3-face. If a
configuration of a set is not compatible with other configurations
of the set (i.e., the configuration cannot enters any ground-state
structure in this 3-face), then it is separated by the symbol |.
The character of transition between two regions (phases) depends
on the ground-state structures at their boundary (i.e., at their
common 3-face). If the set of configurations for a 3-face
generates only the structures which are the ground-state ones for
the full-dimensional regions sharing this 3-face and nothing else,
then there is a first-order phase transition between these regions
(notation ``Jump'' in Table~III). This kind of phase transitions
exists between phases 1 and $\bar 1$, 1 and 3, 1 and 9, 3 and 9,
as well as 4 and 10 (and also between pairs of symmetric phases).
If for any value of magnetization from the interval between the
values of magnetization for two neighboring full-dimensional
phases, at least one ground-state structure can be constructed,
then the transition between these phases is continuous (notation
``Cont.'' in Table~III). As it is clear from Table~III, the most
part of transitions just possess this property. The situation
between phases 4 and 9 is more complicated: there are both
discontinuous and continuous phase transitions between them.

An interesting and important question concerns the disorder and
entropy at the boundaries of full-dimensional regions, in
particular, in the 3-faces. Let us consider, for instance, a
3-face between the regions 1 and 10. A typical example of a
structure in this 3-face is shown in Fig.~9(a): the structure
consists of diagonal spin-down chains (along $J_3$ bonds)
separated by diagonal spin-up chains whose number can be arbitrary
odd but not less than three. This is a simple mixture of
structures 1 and 10. It is clear that order exists along these
chains but there is disorder in the perpendicular direction. Such
a disorder can be called one-dimensional. It does not lead to
macroscopic degeneracy: the entropy per site tends to zero if
dimensions of the lattice tend to infinity. A similar simple
mixtures of full-dimensional structures exist also at boundaries
of phases 1 and 5, 1 and 8, 3 and 4, 3 and 8, as well as 3 and 10.
At the boundary between phases 9 and 10, the ground-state
structures are not a simple mixture of full-dimensional structures
9 and 10 but rather a kind of their hybrid, although the disorder
is one-dimensional there. They look like structure 10 with
additional spin-down diagonal chains normal to the similar chains
of structure 10 [Fig.~9(b)].

\begin{figure*}[]
\begin{center}
\includegraphics[scale = 0.85]{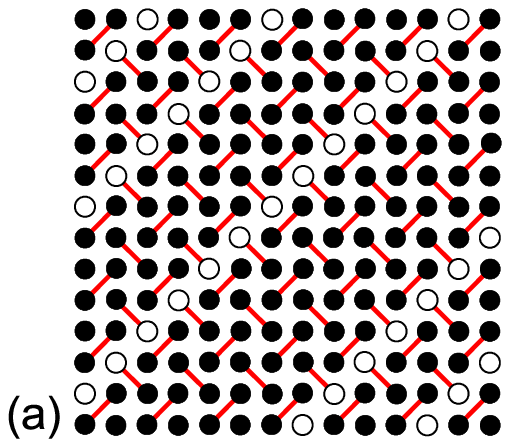}
\hspace{0.5cm}
\includegraphics[scale = 0.85]{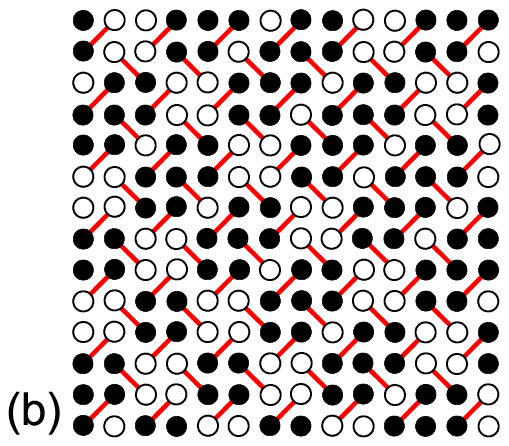}
\hspace{0.5cm}
\includegraphics[scale = 0.85]{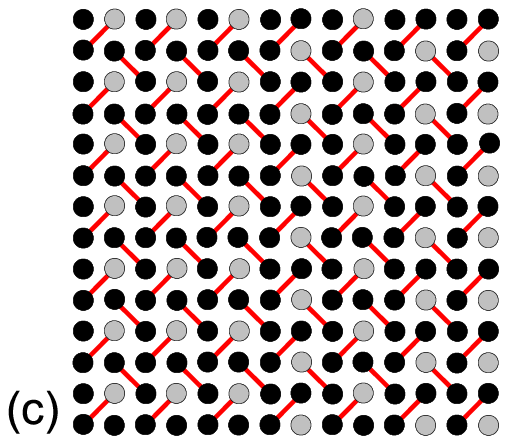}
\caption{Examples of the ground-state structures at the boundaries
between phases (a) 1 and 10 as well as (b) 9 and 10. The disorder
at theses boundaries is one-dimensional. (c) An example of
ground-state structures at the boundary between phases 1 and 6.
Each gray circle can be either open or solid, hence, the disorder
is two-dimensional at this boundary. If all the gray circles are
open, then we obtain a mixture of structures $6a$ and $6b$, which
is a ground-state structure in region 6.}
\label{fig9}
\end{center}
\end{figure*}

Another kind of disorder occurs, for instance, at the boundary
between phases 1 and 6. The ground-state structures at this
boundary can be obtained from a mixture of structures $6a$ and
$6b$ (which is a ground state in region 6) by flipping a part of
spins down [Fig.~9(c)]. It is clear that this disorder is
two-dimensional since infinite number of local changes can be made
in the structures within the bounds of the ground state. Every
fifth spin in the structure depicted in Fig.~9(c) is ``free,''
that is, can be directed downward as well as upward. Hence, the
entropy per site is equal to $\frac15 \ln 2$ or maybe more, since
Fig.~9(b) does not exhaust all structures.

\begin{figure}[]
\begin{center}
\includegraphics[scale = 0.85]{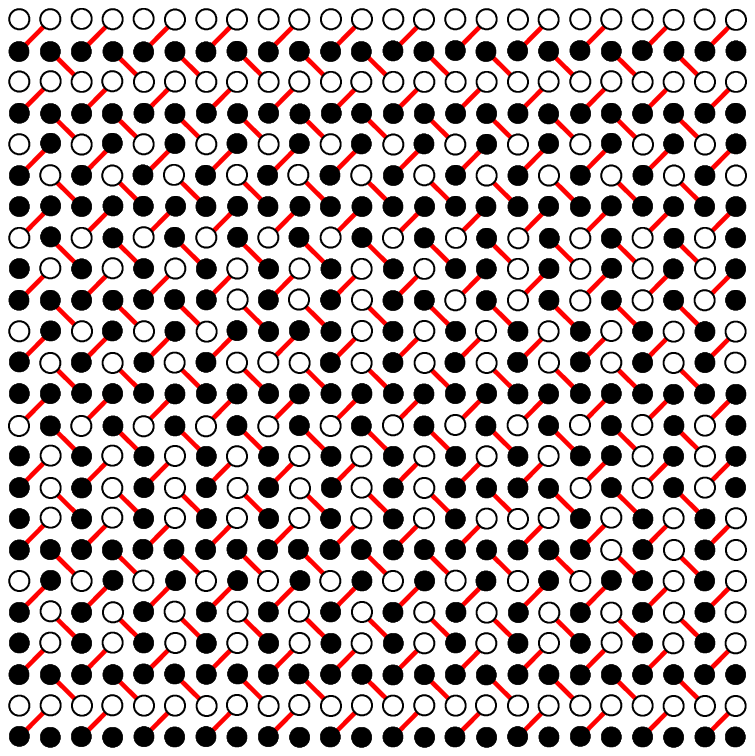}
\caption{An example of a ground-state structure at the boundary
between phases 4 and 5. It seems that the disorder is
one-dimensional, however we cannot prove it as yet.}
\label{fig10}
\end{center}
\end{figure}

\begin{figure}[]
\begin{center}
\includegraphics[scale = 0.85]{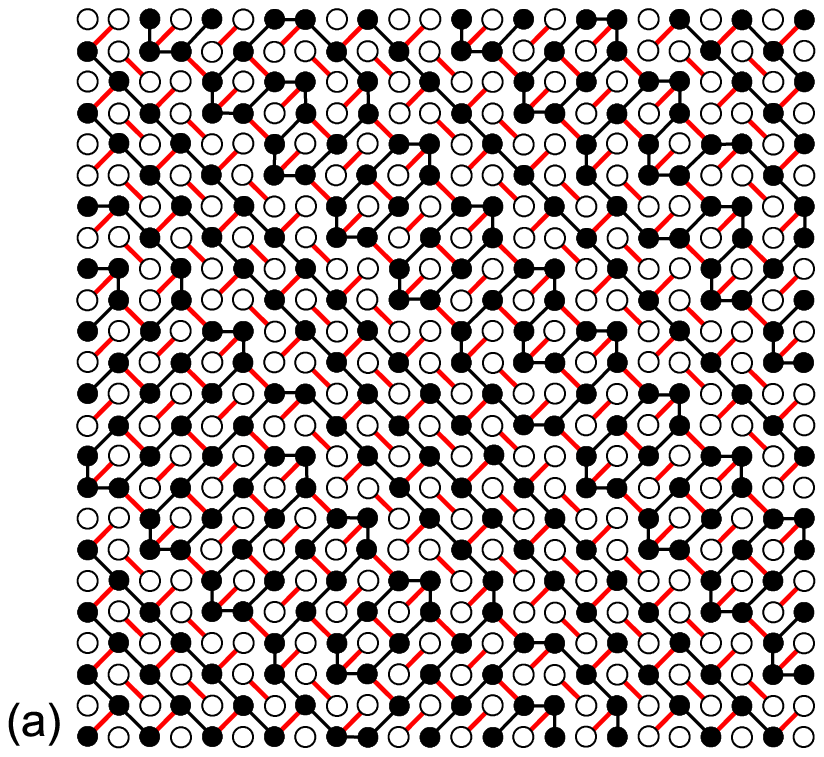}

\vspace{0.25cm}

\includegraphics[scale = 0.85]{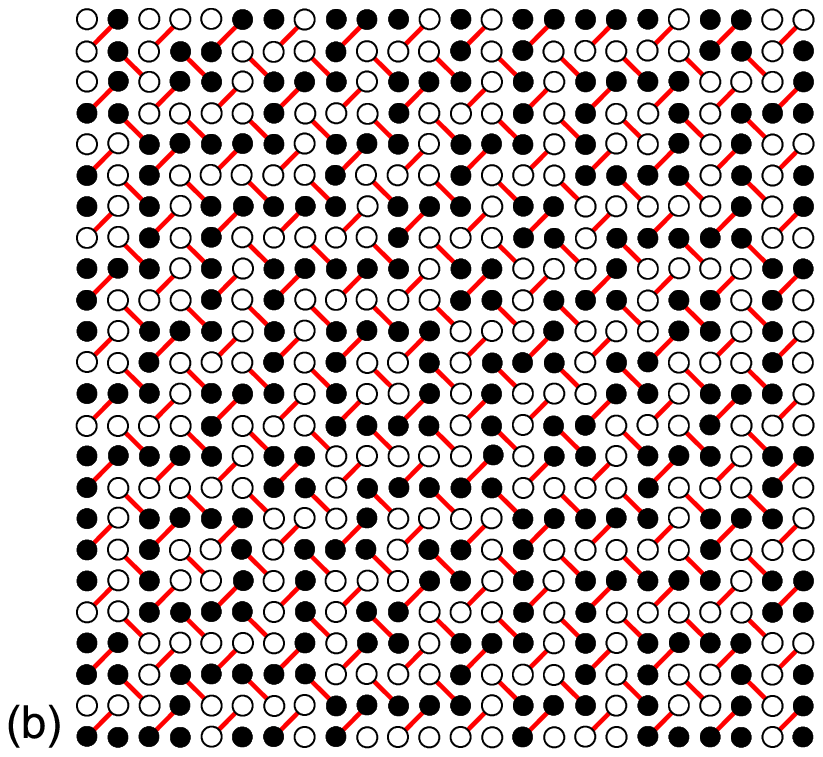}
\caption{Examples of ground-state structures at the boundary
between phases (a) 3 and 5 as well as (b) 5 and 8.  Macroscopic
degeneracy (residual entropy) occurs only at the boundary between
phases 3 and 5 but not at the boundary between phases 5 and 8
although in the latter case the disorder exists in two
perpendicular directions independently. Some sites are connected
with lines for better visualization.}
\label{fig11}
\end{center}
\end{figure}

In Table~III, the dimensionality of disorder is indicated in the
fourth column. To find it is sometimes not so easy as in the cases
considered above. For instance, we could not determine the
dimensionality of disorder at the boundary between phases 4 and 5
(see Fig.~10). The disorder in this 3-face is most likely
one-dimensional but it still should be proved. Let us analyze the
disorder at some other 3-faces where such analysis is nontrivial.

One can see from Fig.~11 that at the boundary between phases 3 and
5 as well as 5 and 8 the disorder exists in two perpendicular
directions independently. However, at the first boundary the
disorder is two-dimensional (and the entropy per site is nonzero)
and at the second one the disorder is one-dimensional. The
structures at the boundary between phases 5 and 8 are interesting
in the sense that they are determined by a single condition: all
the ``empty'' squares should have configuration \usebox{\uudd}.

An interesting set of structures also occurs at the boundary
between phases 4 and 9. The disorder at this boundary is
one-dimensional, since the structure is completely determined by a
zigzag-stripe an example of which is depicted in Fig.~12. The
structure closest to structure 9 in this set is an ordered
structure with $m/m_s = 1/5$ [Fig.~13(a)]. Hence, there is a jump
between structure 9 and this structure and then a continuous
transition to phase 4, that is, at the boundary between phases 4
and 9 there occur both a jump and a continuous transition. In
addition to the structures of the type shown in Fig.~12, there is
one structure of another type at the boundary between phases 4 and
9. This is a structure with $m/m_s = 1/4$ shown in Fig.~13(b). All
these structures (except for structure 4) are chiral (see Fig.~8
and its explanation in the text).

Another set of structures, where it is difficult to determine the
dimensionality of disorder, occurs at the boundary between phases
4 and 6. An example of a structure at this boundary is shown in
Fig.~14. It seems that the disorder is two-dimensional, however, a
more profound analysis shows that the disorder is nevertheless
one-dimensional since, as in phase 6, the structural element shown
in Fig.~7 generates an infinite half-stripe.

At the boundary between phases 3 and 4 a collection of stripe
structures exists. Most probably, some of these give rise to
fractional magnetization plateaus in TmB$_4$. We do not describe
these structures here, since this is done in
Ref.~\onlinecite{bib9}. The interaction $J_3$ does not lift the
degeneracy at this boundary.

\begin{figure}[]
\begin{center}
\includegraphics[scale = 0.85]{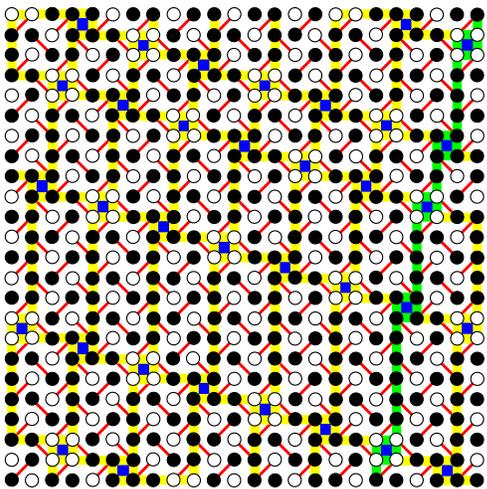}
\caption{An example of ground-state structures at the boundary
between phases 4 and 9. The rectangles formed by the ferromagnetic
chains together with ``open'' and ``solid'' squares are shown as
well as one zigzag-stripe that generates the whole structure. The
structure is chiral: the SS bonds twist clockwise around ``open''
squares and counterclockwise around ``solid'' ones (and vice versa
for the chiral twin).}
\label{fig12}
\end{center}
\end{figure}

\begin{figure}[]
\begin{center}
\includegraphics[scale = 0.85]{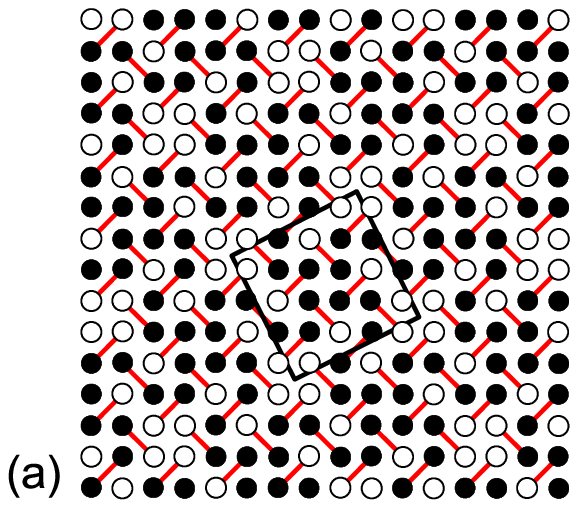}

\vspace{0.25cm}

\includegraphics[scale = 0.85]{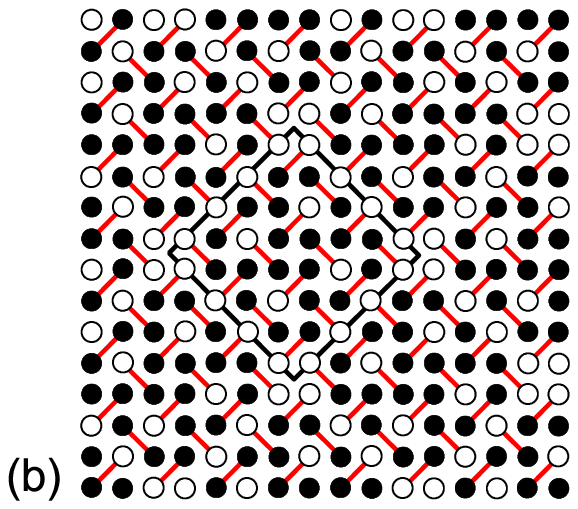}
\caption{Examples of ground-state structures at the boundary
between phases 4 and 9: (a) $m/m_s = 1/5$, (b) $m/m_s = 1/4$. Unit
cells are indicated.}
\label{fig13}
\end{center}
\end{figure}

\begin{figure}[]
\begin{center}
\includegraphics[scale = 0.85]{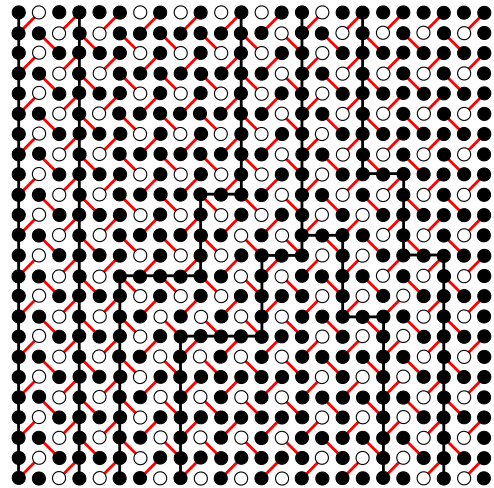}
\caption{An example of a ground-state structure at the boundary
between phases 4 and 6. Lines separate domains of phases 4 and 6.}
\label{fig14}
\end{center}
\end{figure}

\section{Ground-state structures at the 2-faces of full-dimensional regions}

Although the ground states in the 2-faces of full-dimensional
regions are not so important as the ground states in the 3-faces,
let us consider them for completeness. We denote a 2-face by two
basic vectors (which generate it) in curly brackets. In Table~III,
vectors of each 3-face are ordered in the way that each pair of
neighboring vectors (the first and the last are neighbors as well)
generates a 2-face. The set of ground-state configurations for a
2-face is an intersection of the basic sets of configurations for
these vectors. The 2-faces of full-dimensional regions for $h
\geqslant 0$ and the ground-state configurations for them are
presented in Table~IV. The third, fourth, fifth, and sixth columns
in Table~IV indicate, respectively, the dimensionality of disorder
in a 2-face, the full-dimensional regions sharing this 2-face, the
coordinates of the point that corresponds to this 2-face in the
plane $(h, J_2)$, and the conditions for the existence of such a
point in this plane (if the coordinates are not indicated in
parenthesis, then the 2-face does not intersect the plane $(h,
J_2)$ or lies completely in it for some values of $J_1$ and
$J_3$).

\begin{table*}
\caption{Basic rays and ground-state configurations for the
2-faces of the full-dimensional regions for the Ising model on the
extended SS lattice ($h \geqslant 0$).}
\begin{ruledtabular}
\begin{tabular}{llcclc}
\multicolumn{1}{c}{}&\multicolumn{1}{c}{Ground-state configurations}&&\multicolumn{1}{c}{Full-dimensional}&\multicolumn{1}{c}{Coordinates}&\multicolumn{1}{c}{Conditions for existence}\\
\multicolumn{1}{c}{2-face}&\multicolumn{1}{c}{for the 2-face}&Disorder&\multicolumn{1}{c}{structures}&\multicolumn{1}{c}{in the plane $(h, J_2)$}&\multicolumn{1}{c}{in the plane $(h, J_2)$}\\
\hline\\[-2mm]
$\{\mathbf{r}_1, \mathbf{r}_2\}$&\usebox{\sdddd} \usebox{\suddu} \usebox{\sduud} \usebox{\suuuu} $|$ \usebox{\sdudd} \usebox{\suduu} $\parallel$ \usebox{\dddd} \usebox{\uddu} \usebox{\uuuu}&0&1, $\bar 1$, 3&$h = 0$, $J_1 = 0$&$J_2 < 0$, $J_3 < 0$\\
$\{\mathbf{r}_1, \mathbf{r}_4\}$&\usebox{\sdddd} \usebox{\suudd} \usebox{\suuuu} $|$ \usebox{\suddd} \usebox{\sduuu} $\parallel$ \usebox{\dddd} \usebox{\uddu} \usebox{\uuuu}&0&1, $\bar 1$, 9&$(0, -2 J_1)$&$J_1 < 0$, $J_3 < 0$\\
$\{\mathbf{r}_1, \mathbf{r}_7\}$&\usebox{\suddu} \usebox{\sduud} \usebox{\sduuu} \usebox{\suduu} \usebox{\suuuu} $\parallel$ \usebox{\dddd} \usebox{\uddu} \usebox{\uuuu}&1&1, 3, 10&$(4J_1, 0)$&$J_1 > 0$, $J_3 < 0$\\
$\{\mathbf{r}_1, \mathbf{r}_8\}$&\usebox{\suudd} \usebox{\sduuu} \usebox{\suduu} \usebox{\suuuu} $|$ \usebox{\suddd} \usebox{\suddu} $\parallel$ \usebox{\dddd} \usebox{\uddu} \usebox{\uuuu}&1&1, 9, 10&$h = J_2$, $J_1 = 0$&$J_2 > 0$, $J_3 < 0$\\
$\{\mathbf{r}_1, \mathbf{r}_{10}\}$&\usebox{\suudd} \usebox{\suddu} \usebox{\sduud} \usebox{\sduuu} \usebox{\suduu} $\parallel$ \usebox{\dddd} \usebox{\uddu} \usebox{\uuuu}&1&3, 9, 10&$(2J_1, 2J_1)$&$J_3 < -J_1 < 0$\\

$\{\mathbf{r}_2, \mathbf{r}_3\}$&\usebox{\sdddd} \usebox{\sdudd} \usebox{\suddu} \usebox{\sduud} \usebox{\suduu} \usebox{\suuuu} $\parallel$ \usebox{\dddd} \usebox{\uudd} \usebox{\uuuu}&1&1, $\bar 1$, 8&$h = 0$, $J_3 = -J_1$&$J_1 < 0$, $J_2 < 0$\\
$\{\mathbf{r}_2, \mathbf{r}_6\}$&\usebox{\sdddd} \usebox{\sdudd} \usebox{\suddu} \usebox{\sduud} \usebox{\suduu} \usebox{\suuuu} $\parallel$ \usebox{\uudd} \usebox{\duuu} \usebox{\uuuu}&2&1, 7, 8&$h = 2J_3$, $J_1 = 0$&$J_2 < 0$, $J_3 > 0$\\
$\{\mathbf{r}_2, \mathbf{r}_7\}$&\usebox{\suddu} \usebox{\sduud} \usebox{\suduu} \usebox{\suuuu}&2&1, 3, 7&$h = 4J_1$, $J_3 = 0$&$J_1 > 0$, $J_2 < 0$\\
$\{\mathbf{r}_2, \mathbf{r}_9\}$&\usebox{\sdddd} \usebox{\sdudd} \usebox{\suddu} \usebox{\sduud} \usebox{\suduu} \usebox{\suuuu} $\parallel$ \usebox{\uudd} \usebox{\uddu} \usebox{\duuu}&2&3, 7, 8&$h = 2J_1$, $J_3 = J_1$&$J_1 > 0$, $J_2 < 0$\\

$\{\mathbf{r}_3, \mathbf{r}_4\}$&\usebox{\sdddd} \usebox{\suddd} \usebox{\suudd} \usebox{\sduuu} \usebox{\suuuu} $\parallel$ \usebox{\dddd} \usebox{\uudd} \usebox{\uuuu}&2&1, $\bar 1$, 5&$(0, -2J_1 - 2J_3)$&$0 < J_3 < -J_1$\\
$\{\mathbf{r}_3, \mathbf{r}_6\}$&\usebox{\uudd} \usebox{\uuuu}&2&1, 5, 8&$(2J_1 + 2J_3, 0)$&$0 < -J_1 < J_3$\\

$\{\mathbf{r}_4, \mathbf{r}_8\}$&\usebox{\suddd} \usebox{\suudd} \usebox{\sduuu} \usebox{\suuuu}&2&1, 5, 9&$h = J_2 + 2 J_1$, $J_3 = 0$&$-J_2 < 2J_1 < 0$, $J_2 > 0$\\

$\{\mathbf{r}_5, \mathbf{r}_8\}$&\usebox{\suddd} \usebox{\suudd} \usebox{\suddu}&2&4, 5, 9&$h = J_2 - 2J_1$, $J_3  = 0$&$0 < 2J_1 < J_2$, $J_2 > 0$\\
$\{\mathbf{r}_5, \mathbf{r}_9\}$&\usebox{\suddd} \usebox{\suudd} \usebox{\suddu} \usebox{\sduud} \usebox{\sduuu} $\parallel$ \usebox{\uudd} \usebox{\uddu} \usebox{\duuu}&2&3, 4, 5&$(2J_3, 2J_1 - 2J_3)$&$0 < J_3 < J_1$\\
$\{\mathbf{r}_5, \mathbf{r}_{10}\}$&\usebox{\suudd} \usebox{\suddu} \usebox{\sduud} \usebox{\sduuu}* $\parallel$ \usebox{\dddd} \usebox{\uddu} \usebox{\uuuu} \usebox{\duuu}*
\footnote{\mbox{Configurations marked by asterisk enter the structures in blocks shown in Fig.~4.}}&2&3, 4, 9&$(-2J_3, 2J_1)$&$-J_1 < J_3 < 0$\\

$\{\mathbf{r}_6, \mathbf{r}_7\}$&\usebox{\suddu} \usebox{\sduud} \usebox{\sduuu} \usebox{\suduu} \usebox{\suuuu} $\parallel$ \usebox{\uudd} \usebox{\duuu} \usebox{\uuuu}&2&1, 6, 7&$(4J_1 + 2J_3, 0)$&$J_1 > 0$, $J_3 > 0$\\
$\{\mathbf{r}_6, \mathbf{r}_8\}$&\usebox{\suddd} \usebox{\suudd} \usebox{\suddu} \usebox{\sduuu} \usebox{\suduu} \usebox{\suuuu} $\parallel$ \usebox{\uudd} \usebox{\duuu} \usebox{\uuuu}&2&1, 5, 6&$h = J_2 + 2J_3$, $J_1 = 0$&$J_2 > 0$, $J_3 > 0$\\
$\{\mathbf{r}_6, \mathbf{r}_9\}$&\usebox{\uudd} \usebox{\duuu}&2&5, 6, 7, 8&$(2J_3, 0)$&$0 < J_1 < J_3$\\

$\{\mathbf{r}_7, \mathbf{r}_8\}$&\usebox{\suddu} \usebox{\sduuu} \usebox{\suduu} \usebox{\suuuu}&2&1, 4, 6, 10&$h = J_2 + 4J_1$, $J_3 = 0$&$J_1 > 0$, $J_2 > 0$\\
$\{\mathbf{r}_7, \mathbf{r}_9\}$&\usebox{\suddu} \usebox{\sduud} \usebox{\sduuu} \usebox{\suduu} \usebox{\suuuu} $\parallel$ \usebox{\uddu} \usebox{\duuu} $|$ \usebox{\uudd}&2&3, 4, 6, 7&$(4J_1 - 2J_3, 0)$&$0 < J_3 < J_1$\\
$\{\mathbf{r}_7, \mathbf{r}_{10}\}$&\usebox{\suddu} \usebox{\sduud} \usebox{\sduuu}* \usebox{\suduu}* $\parallel$ \usebox{\uddu} \usebox{\uuuu} \usebox{\duuu}* $|$ \usebox{\dddd}&1&3, 4, 10&$(4J_1 + 2J_3, -2J_3)$&$-J_1 < J_3 < 0$\\

$\{\mathbf{r}_8, \mathbf{r}_9\}$&\usebox{\suddd} \usebox{\suudd} \usebox{\suddu} \usebox{\sduuu} \usebox{\suduu} \usebox{\suuuu} $\parallel$ \usebox{\uudd} \usebox{\uddu} \usebox{\duuu}&2&4, 5, 6&$h = J_2 + 2J_1$, $J_1 = J_3$&$J_1 > 0$, $J_2 > 0$\\
$\{\mathbf{r}_8, \mathbf{r}_{10}\}$&\usebox{\suudd} \usebox{\suddu} \usebox{\sduuu}* \usebox{\suduu}* $\parallel$ \usebox{\dddd} \usebox{\uddu} \usebox{\uuuu} \usebox{\duuu}*&2&4, 9, 10&$h = J_2$, $J_1 = -J_3$&$J_1 > 0$, $J_2 > 0$\\

$\{\mathbf{r}_9, \mathbf{r}_9^{-}\}$&\usebox{\uudd} \usebox{\uddu}&2&3, 5, 8&$J_1 = J_3$, $J_2 = 0$&$J_1 > 0$
\label{table4}
\end{tabular}
\end{ruledtabular}
\end{table*}

It is easy to determine the dimensionality of disorder in the most
part of 2-faces, taking into account that the disorder in a 2-face
cannot be lower than the disorder in a 3-face bounded by this
2-face. This task is difficult, however, for some 2-faces. Let us
consider such 2-faces. In the 2-face $\{\mathbf{r}_3,
\mathbf{r}_4\}$, the disorder is two-dimensional. This is clear
from Fig.~15. It is more difficult to show the disorder in the
2-face $\{\mathbf{r}_5, \mathbf{r}_{10}\}$ is two-dimensional too.
This is illustrated in Fig.~16, where the zigzag-strips going from
left to right can both descend and ascend a little. In the 2-face
$\{\mathbf{r}_8, \mathbf{r}_{10}\}$ the disorder is
two-dimensional. This is clear from Fig.~17, where each pair of
gray circles should contain one open and one solid circle.

Finally, let us note that the disorder is two-dimensional in all
the basic rays.

\begin{figure}[]
\begin{center}
\includegraphics[scale = 0.85]{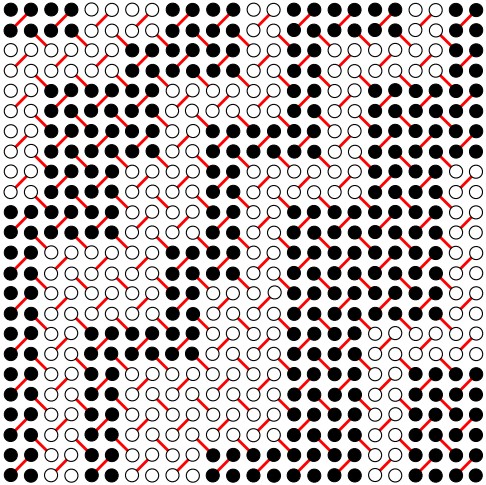}
\caption{An example of a ground-state structure at the boundary
between phases 1, $\bar 1$, and 5 (2-face $\{\mathbf{r}_3,
\mathbf{r}_4\}$), which shows that the disorder at this boundary
is two-dimensional.}
\label{fig15}
\end{center}
\end{figure}

\begin{figure}[]
\begin{center}
\includegraphics[scale = 0.85]{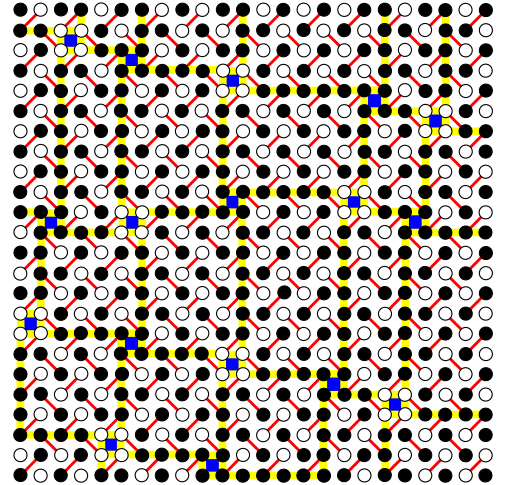}
\caption{Example of a ground-state structure at the boundary
between phases 3, 4, and 9 (2-face $\{\mathbf{r}_5,
\mathbf{r}_{10}\}$). The rectangles formed by the ferromagnetic
chains together with ``open'' and ``solid'' squares are shown.}
\label{fig16}
\end{center}
\end{figure}

\begin{figure}[]
\begin{center}
\includegraphics[scale = 0.85]{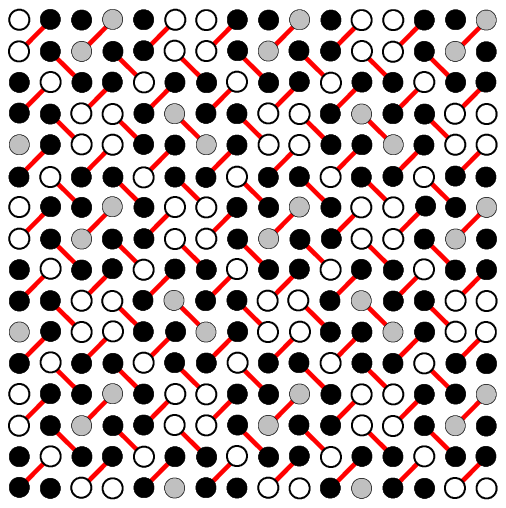}

\vspace{0.25cm}

\includegraphics[scale = 0.85]{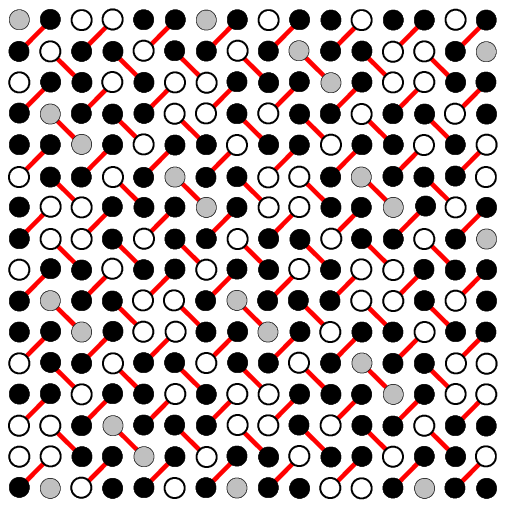}
\caption{Examples of ground-state structures at the boundary
between phases 4, 9, and 10 (2-face $\{\mathbf{r}_8,
\mathbf{r}_{10}\}$), which show that the disorder at this boundary
is two-dimensional. In each pair of gray circles, one circle
should be open and another one --- solid. Both structures can mix
with each other.}
\label{fig17}
\end{center}
\end{figure}

\section{Ground-state phase diagrams and fractional magnetization plateaus}

To make our results more usable, we present all the types of
ground-state phase diagrams in the plane $(h, J_2)$. There are
three types of diagrams if $J_1 < 0$ (Fig.~18) and four types if
$J_1 > 0$ (Fig.~19). In these diagrams, the lines between
neighboring regions correspond to the 3-faces and the points where
three or more regions converge correspond to the 2-faces. If there
is a first-order phase transition between the neighboring phases,
then the line separating the corresponding regions is solid red;
if the transition is continuous, then the line is solid black. The
dash-dotted green line between regions 4 and 9 corresponds to a
jump along with a continuous transition. Regions 6, 7, and 10,
which give rise to an 1/2-plateau, are colored and region 4, which
gives rise to a 1/3-plateau, is shaded.

\begin{figure*}[]
\begin{center}
\includegraphics[scale = 0.8]{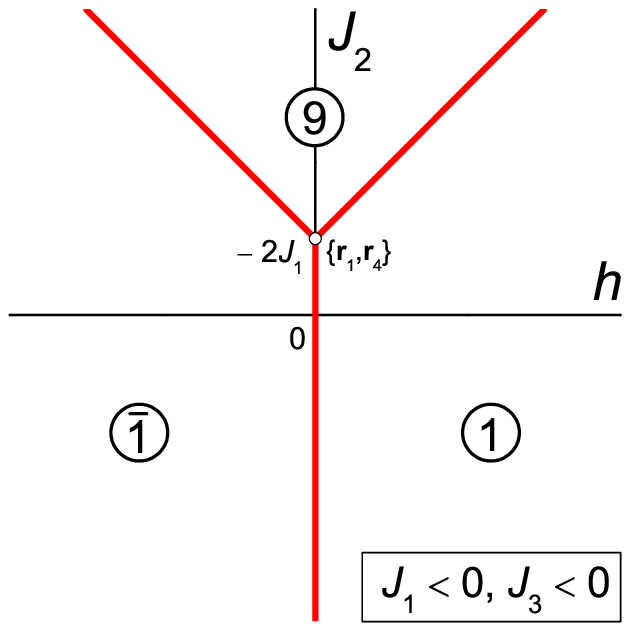}
\hspace{0.5cm}
\includegraphics[scale = 0.8]{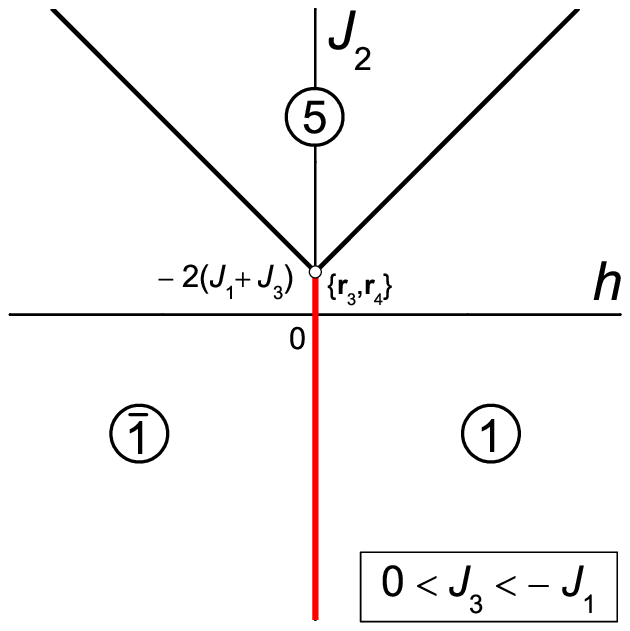}
\hspace{0.5cm}
\includegraphics[scale = 0.8]{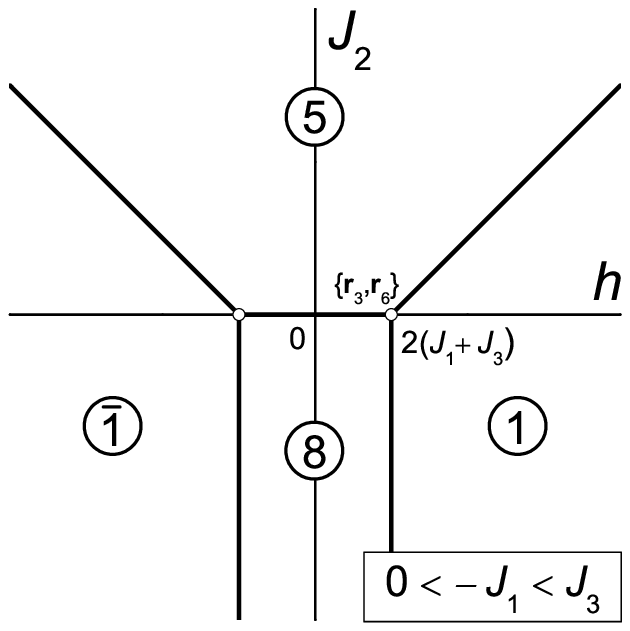}
\caption{Ground-state phase diagrams of the Ising model on the
extended SS lattice for $J_1 < 0$. Black and red lines correspond
to continuous phase transitions and jumps, respectively. (See also
Table~öö and Figs.~5~and~6).}
\label{fig18}
\end{center}
\end{figure*}

\begin{figure*}[]
\begin{center}
\includegraphics[scale = 0.8]{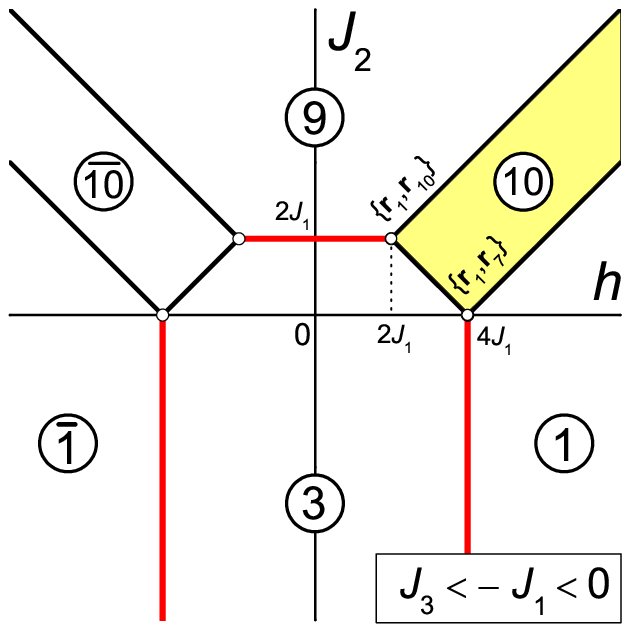}
\hspace{0.5cm}
\includegraphics[scale = 0.8]{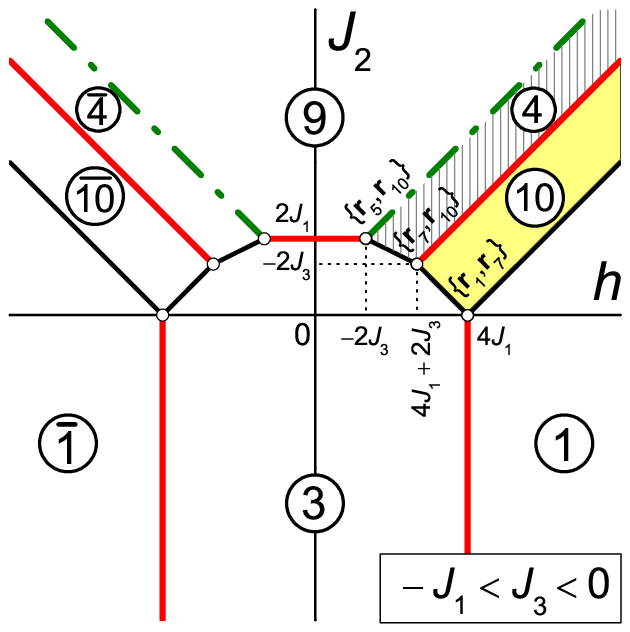}

\vspace{0.5cm}

\includegraphics[scale = 0.8]{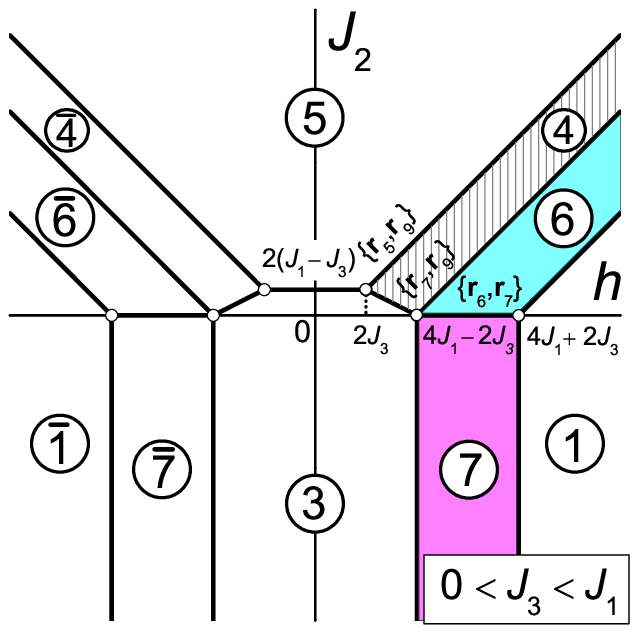}
\hspace{0.5cm}
\includegraphics[scale = 0.8]{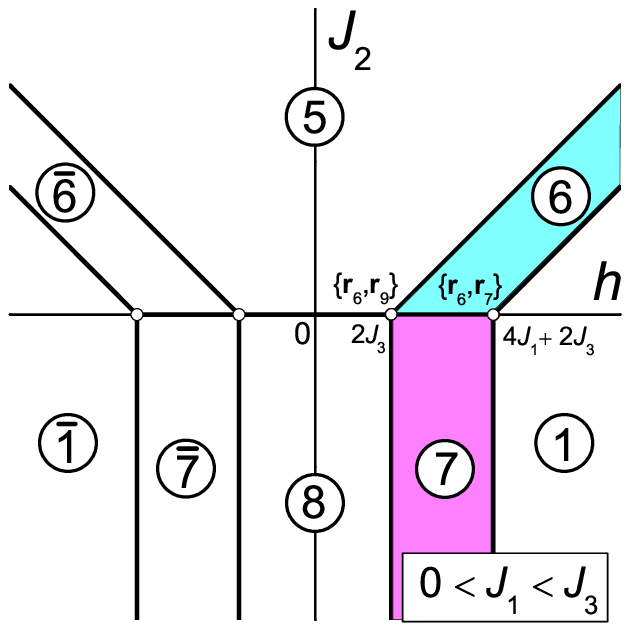}
\caption{Ground-state phase diagrams of the Ising model on the
extended SS lattice for $J_1 > 0$. Black and red lines correspond
to continuous phase transitions and jumps, respectively; a green
dash-dotted line denotes a jump together with a continuous phase
transition. The regions which give rise to an 1/2-plateau are
colored. The region of 1/3-plateau phase is shaded. (See also
Table~öö and Figs.~5~and~6).}
\label{fig19}
\end{center}
\end{figure*}

As one can see from Fig.~19 and Table~II, the width of the
1/2-plateau generated by phase 10 is equal to $2J_2$ and $4J_1$
for sequences of phases 3 -- 10 -- 1 and 9 -- 10 -- 1; for all the
rest of sequences that give rise to this plateau its width is
equal to $|4J_3|$. The widths of zero plateaus and fractional ones
for various sequences of phases as well as conditions for their
existence are presented in Table~V.

\begin{table*}
\caption{The widths of the fractional plateaus for various ways of
transitions from the zero-field phase to the ferromagnetic one.}
\begin{ruledtabular}
\begin{tabular}{ccccc}
\multicolumn{1}{c}{Sequence}&Width&Width&Width&\\
\multicolumn{1}{c}{of phases}&of zero plateau&of 1/3-plateau&of 1/2-plateau&Conditions for existence\\
\hline\\[-2mm]
3 -- 10 -- 1&$4J_1-J_2$&0&$2J_2$&$J_1 > 0$, $J_3 < 0$, $0 < J_2 < \min[2J_1,-2J_3]$\\
3 -- 7 -- 1&$4J_1-2J_3$&0&$4J_3$&$0< J_3 < J_1$\\
5 -- 6 -- 1&$J_2+2J_3$&0&$4J_3$&$0 < J_1 < J_3$\\
8 -- 7 -- 1&$2J_3$&0&$4J_3$&$0 < J_1 < J_3$\\
9 -- 10 -- 1&$J_2$&0&$4J_1$&$J_3 < -J_1 < 0$, $J_2 > 2J_1$\\
3 -- 4 -- 10 -- 1&$4J_1-2J_2-2J_3$&$3J_2+6J_3$&$-4J_3$&$-J_1 < J_3 < 0$, $-2J_3 < J_2 < 2J_1$\\
9 -- 4 -- 10 -- 1&$-2J_1+J_2-2J_3$&$6(J_1+J_3)$&$-4J_3$&$-J_1 < J_3 < 0$, $J_2 > 2J_1$\\
3 -- 4 -- 6 -- 1&$4J_1-2J_2-2J_3$&$3J_2$&$4J_3$&$0<J_3<J_1$, $0 < J_2 < 2(J_1 - J_3)$\\
5 -- 4 -- 6 -- 1&$-2J_1+J_2+4J_3$&$6(J_1-J_3)$&$4J_3$&$0 < J_3 < J_1$, $J_2 > 2(J_1 - J_3)$
\label{table5}
\end{tabular}
\end{ruledtabular}
\end{table*}

Which sequence of phases corresponds to the magnetization process
in ErB$_4$? A single 1/2-plateau has been observed in this
compound, therefore, only five initial sequences in Table~V are
possible. However, since interactions $J_1$ and $J_2$ are
antiferromagnetic and approximately equal and interaction $J_3$ is
ferromagnetic and relatively large, \cite{bib5} only the sequence
3 -- 10 -- 1 remains. This contradicts the statement of
Ref.~\onlinecite{bib5} about the magnetic structure of ErB$_4$ for
zero field. According to this reference, there should be structure
AF1 rather than the N\'{e}el phase (AF3). The structure AF1, in
the ground-state phase diagram ($J_1 = J_2$, $h = 0$) presented in
this reference, should correspond to our structure 5, however,
some other structure is depicted there; maybe this is a simple
inadvertence.

On the basis of the ground-state structures that we have found
previously, \cite{bib9} the authors of Ref.~\onlinecite{bib18}
have constructed a ground-state phase diagram for $J_1 = J_2 > 0$
and for arbitrary $J_3$ and then obtained numerically a spin
supersolid phase in the quantum model with strong Ising
anisotropy. They argue that this ground state exists in ErB$_4$
which, without magnetic field, has magnetic structure 5. However,
structure 5 is possible only under the condition $J_3 > J_1
> 0$ which looks unrealistic. It seems that experimental results
in favor of this structure in ErB$_4$ without magnetic field
\cite{bib6,bib19} and the experimental results yielding that
interaction $J_3$ is ferromagnetic and large \cite{bib5} are
contradictory.

The structure with $m/m_s = 1/2$, presented in
Ref.~\onlinecite{bib4} for TmB$_4$ is erroneous as well. It is a
mixture of structures 1 and 5. In the model under consideration,
this structure exists at the boundary between phases 1 and 5 where
$J_1 < 0$. The authors of Refs.~\onlinecite{bib4} and
\onlinecite{bib5} state that, in TmB$_4$, interaction $J_1 \approx
J_2$ is antiferromagnetic, therefore, it is unlikely that this
structure could occur in this compound.

It is also unlikely that, in TmB$_4$, the structure (3,7)$_4$ with
$m/m_s = 1/9$ presented in Ref.~\onlinecite{bib4} (Fig.~20) could
exist. In comparison with the structure (3,4)$_4$ that we proposed
in Ref.~\onlinecite{bib4}, all the antiferromagnetic chains in
this structure are shifted. In the model under consideration, the
structure (3,7)$_4$ exists at the boundary between phases 3 and 7,
that is, for positive values of $J_1$ and $J_3$ but negative
values of $J_2$.

In Ref.~\onlinecite{bib5} the following experimental values of
interaction parameters for TmB$_4$ are presented: $J_1 = J_2 =
0.85~K$ and $J_3 = 0.3~K$. These correspond to the sequence of
phases 3 -- 4 -- 6 -- 1, however, then the 1/3-plateau is
approximately twice as wide as the 1/2-plateau. [If $J_3$ were
ferromagnetic ($J_3 = -0.3~K$), then the 1/2-plateau would be
twice as wide].  The 1/3-plateau has not been observed in TmB$_4$.
Maybe the longer-range interactions, giving rise to the fractional
plateaus 1/6, 1/7..., remove at the same time the 1/3-plateau.

In Ref.~\onlinecite{bib17}, to explain the appearance of
fractional magnetization plateaus in TmB$_4$, a spin-electron
model has been considered. For the 1/2-plateau the structure $6b$
(or symmetric one) has been obtained. The SS diagonals are not
shown in the figures of Ref.~\onlinecite{bib17}, therefore we do
not know whether the structures obtained there are of the same
type as in Ref.~\onlinecite{bib4} or as in Ref.~\onlinecite{bib9}
(Fig.~20). It is easy to show that the interaction $J_4$ (the
next-nearest neighbors along edges) lifts the degeneracy in phase
6 and stabilizes one of the structures $6a$ or $6b$. The
ferromagnetic (antiferromagnetic) interaction $J_4$ stabilizes the
structure $6a$ \cite{bib10,bib11} ($6b$).

\begin{figure}[]
\begin{center}
\includegraphics[scale = 0.85]{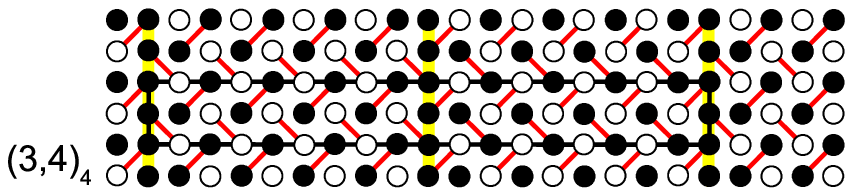}

\vspace{0.5cm}

\includegraphics[scale = 0.85]{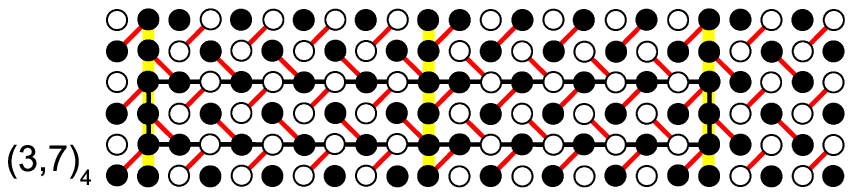}
\caption{Structure (3,4)$_4$ for the 1/9-plateau in TmB$_4$,
proposed in Ref.~\onlinecite{bib9}, and structure (3,7)$_4$,
proposed for this plateau in Ref.~\onlinecite{bib4}. These two
structures differ by the positions of the antiferromagnetic
chains. Unit cells are indicated.}
\label{fig20}
\end{center}
\end{figure}

It is interesting that in the case where all the three
interactions are antiferromagnetic ($J_1 > 0$, $J_2 > 0$, and $J_3
> 0$) the full-dimensional phases for $h \geqslant 0$ (1, 3, 4, 5, and
6) can be considered as one-dimensional ordering of ferro- and
antiferromagnetic chains. A question arises: In order to study the
effect of longer-range interactions on ground states, is it
possible to consider a one-dimensional model with effective
interactions between chains instead of the two-dimensional one?
\cite{bib20}

\section{Summary}

To conclude, we have determined a complete solution of the
ground-state problem for an Ising model on the extended SS lattice
with an additional interaction $J_3$ along the diagonals of
``empty'' squares. We have used the method of basic vectors and
basic sets of cluster configurations that was proposed in our
previous works. Here, however, we generalize the method and
consider configurations of two clusters at once. We have
constructed the ground-state phase diagrams and studied the
ground-state structures in the full-dimensional regions as well as
at their three- and two-dimensional boundaries. This made it
possible to establish that the additional interaction $J_3$ gives
rise to an 1/2-plateau. This plateau can correspond to three
different phases (depending on the interaction parameters), two of
which are partially disordered. In addition to the 1/2-plateau,
another fractional plateau is possible---with the magnetization
1/3. As we have shown earlier, \cite{bib9} this is a single
fractional plateau in the Ising model on the conventional SS
lattice. For some relations between the interaction parameters,
the 1/2-plateau can be a single fractional plateau in the Ising
model on the extended SS lattice as well. Hence, it might be
reasonable to believe that we have explained the origin of a
single fractional plateau---with $m/m_s = 1/2$---in ErB$_4$. As
to TmB$_4$, where no 1/3-plateau but an 1/2-plateau and a sequence
of other fractional plateaus were observed, the theoretical
explanation of the magnetization curve in this compound requires
to study the effect of longer-range interactions. However, the
knowledge of the ground-states at the boundaries of the
full-dimensional regions of the model under consideration makes it
possible to draw some conclusions about the effect of such
interactions.

The advantages of the analysis of the ground states at the
boundaries of full-dimensional regions become obvious when the
results of this paper are compared to the results of the previous
one. \cite{bib9} Having constructed the ground-state phase diagram
for the Ising model on the conventional SS lattice and having
analyzed which of the ground states at boundaries of
full-dimensional regions become full-dimensional if a small
additional interaction $J_3$ is switched on, we have obtained all
the full-dimensional ground-state structures of the Ising model on
the extended SS lattice except for structure 8 (since it occurs
under the condition $J_3 > |J_1|$).

The results obtained here and in Ref.~\onlinecite{bib9} can be
useful for the numerical studies of the origin of the fractional
magnetization plateaus in SS magnets (with Ising anisotropy) for
nonzero temperatures. \cite{bib18}

In the present paper, a development of analytical methods for the
determination of ground states of lattice-gas models and
equivalent spin models is presented. These methods can be used to
investigate the structure of substitutional alloys, that is a very
important problem (a recent initiative of US President Barack
Obama \cite{bib21} gives evidence of this). In our future paper we
will develop the method and show how to study the effect of
longer-range interactions.

\section{Acknowledgments}

The author is grateful to T. Verkholyak and I. Stasyuk for useful
discussions and suggestions and to O. Kocherga for correction of
the text.

\end{document}